\documentclass[fleqn,usenatbib]{mnras}

\usepackage{newtxtext,newtxmath}
\usepackage[T1]{fontenc}
\usepackage{ae,aecompl}

\usepackage{graphicx}	% Including figure files
\usepackage{amsmath}	% Advanced maths commands
\usepackage{amssymb}	% Extra maths symbols
\usepackage{array}
\usepackage{multirow}

\usepackage[dvipsnames]{xcolor}
\colorlet{orange}{green!10!orange!90!}

\usepackage[normalem]{ulem} %for strikeout /sout

\newcommand{\sunpc}{\,\text{M}_{\odot} \, \text{pc}^{-2}}	% solar masses per parsec-squared
\title[High-resolution extinction technique]{A high-resolution extinction mapping technique for face-on disc galaxies}

\author[H. Faustino Vieira et al.]{Helena Faustino Vieira$^{1}$\thanks{E-mail: faustinovieirah@cardiff.ac.uk},
Ana Duarte-Cabral$^{1}$,
Timothy A. Davis$^{1}$,
Nicolas Peretto$^{1}$,
Matthew W. L. Smith$^{1}$,
\newauthor
Miguel Querejeta$^{2}$,
Dario Colombo$^{3}$,
Michael Anderson$^{1}$.
\\
$^{1}$ Cardiff Hub for Astrophysics Research and Technology (CHART), School of Physics \& Astronomy, Cardiff University, The Parade, CF24 3AA Cardiff, UK\\
$^{2}$ Observatorio Astronómico Nacional (IGN), C/ Alfonso XII 3, E28014 Madrid, Spain \\
$^{3}$ Max Planck Institute for Radioastronomy, Auf dem Hügel 69, D-53121 Bonn, Germany \\
}

\date{Accepted XXX. Received YYY; in original form ZZZ}

\pubyear{2023}

\begin{document}
\label{firstpage}
\pagerange{\pageref{firstpage}--\pageref{lastpage}}
\maketitle

\begin{abstract}

We present a new dust extinction technique with which we are able to retrieve parsec-scale gas surface density maps for entire nearby galaxies. The method measures the dust attenuation in optical bands on a pixel-by-pixel basis against a smoothed, reconstructed stellar distribution. The contribution of foreground light along the line-of-sight is calibrated using dust emission observations, assuming that the dust sits in a layer close to the mid-plane of the face-on galaxy. Here, we apply this technique to M51 (NGC 5194) as a proof-of-concept, obtaining a resolution of $0.14"$ (5 pc). Our dust (and gas) surface density map is consistent with independent dust- and CO-based studies at lower resolution. We find that discrepancies between our estimates of surface density and other studies stem primarily from the choice of dust model (i.e. different dust absorption coefficients). When assuming the same dust opacity law, our technique produces surface densities that are consistent with independent studies. This dust extinction technique provides us with gas surface density maps at an unprecedented resolution for full disc coverage studies of nearby galaxies. The resulting well-resolved spatial information opens the possibility for more in-depth examination of the influence of large-scale dynamics (and also stellar feedback mechanisms) on the interstellar medium at parsec-scales, and consequently star formation in nearby galaxies.

\end{abstract}

\begin{keywords}
methods: observational -- dust, extinction -- galaxies: individual (M51) -- galaxies: star formation -- galaxies: ISM -- galaxies: spiral
\end{keywords}

%%%%%%%%%%%%%%%%%%%%%%%%%%%%%%%%%%%%%%%%%%%%%%%%%%

%%%%%%%%%%%%%%%%% BODY OF PAPER %%%%%%%%%%%%%%%%%%

\section{Introduction}

A key process in a galaxy's evolution is its ability to form stars. Star formation (SF) originates in the molecular phase of the interstellar medium (ISM) in galaxies \citep[e.g.][]{young_molecular_1991, bigiel_star_2008}. To better understand the initial conditions of SF it is therefore important to study the molecular gas content in galaxies. By far, the most common method employed in molecular gas studies (since its main component, H$_2$, cannot be directly observed at the cold temperatures of this gas) is CO emission observations \citep[e.g.][]{koda_dynamically_2009,schinnerer_pdbi_2013,schuller_sedigism_2020,leroy_phangs-alma_2021}. With the advancement of instrumentation, interferometric observations are now able to distinguish and resolve giant molecular clouds (GMCs) in nearby galaxies, catapulting us into an exciting era of SF and ISM studies. Still, CO is not a perfect tracer; CO surveys are only sensitive to the "CO-bright" molecular gas since at low column densities CO is not abundant enough to self-shield (unlike H$_2$, the main component of the molecular ISM) and is rapidly photodissociated by the interstellar radiation field \citep[e.g.][]{lada_near-infrared_2007, bolatto_co--h2_2013, penaloza_using_2017, galliano_nearby_2022}. Furthermore, $^{12}$CO is generally optically thick and therefore does not trace the full column density of the gas \citep[e.g.][]{bolatto_co--h2_2013}.

Another proxy for the invisible H$_2$ is interstellar dust. Assuming that dust and gas are well-mixed - a fair assumption given that dust grains act as a catalyst for the otherwise inefficient formation of H$_2$ \citep[e.g.][]{galliano_nearby_2022} - we can use measurements of dust emission or extinction to retrieve the molecular gas content of a galaxy since dust absorbs the stellar light in the visible-UV range and re-emits it at infrared (IR) to sub-millimetre (sub-mm) wavelengths. This, of course, also assumes that the dust-to-gas mass ratio is constant and known. A constant dust-to-gas mass ratio of 0.01 is an often made assumption for the MW and galaxies with similar metallicities \citep[e.g.][]{hildebrand_determination_1983, lada_near-infrared_2007, mentuch_cooper_spatially_2012, de_looze_jingle_2020}, but might be underestimated in different density regimes such as the outskirts of galaxies \citep{smith_far-reaching_2016, giannetti_galactocentric_2017}. Given the long wavelengths involved, as well as the atmospheric filtering that occurs at this range, dust observations are best carried out from space. This, of course, limits the spatial resolution achieved given the physical constraints on instrumentation. Still, dust can be a more reliable tracer of the molecular ISM than CO, and dust extinction in particular is a very promising tracer considering that it is independent of the assumed dust temperature, unlike dust emission \citep[e.g.][]{lada_near-infrared_2007}. Currently, some of the observational techniques used to measure dust extinction are star counts \citep[extinction is measured through comparison of the number of stars of similar brightness within separate regions of a galaxy, e.g.][]{dobashi_atlas_2005} and stellar reddening or colour-excess \citep[where extinction measurements are retrieved by computing the colour difference between individual stars of similar spectral types, e.g.][]{lada_near-infrared_2007,kahre_extinction_2018,barrera-ballesteros_edge-califa_2020}. Both methods rely on being able to resolve individual stars positioned beneath obscuring dust, which is increasingly difficult the further the distance to a galaxy (i.e. the poorer the spatial resolution) and in saturated/high extinction areas. Another significant limitation is that in crowded regions of a galaxy, such as the centre and spiral arms, several stars can blend within a resolution element and consequently a reliable determination of extinction is not possible. Additionally, for the stellar reddening technique, it is necessary to disentangle the true colour of a star from foreground reddening, which will introduce further uncertainties. The dust extinction technique introduced in this paper supersedes the aforementioned methods in terms of the final resolution achieved and is not reliant on resolving individual stars and accurately measuring their colour.

Here, we present a dust extinction technique that allows us to retrieve sub-arcsecond dust (and by extension gas) column density maps for entire galaxies. This method is adapted from dust extinction studies done for the MW in the IR \citep[e.g.][]{bacmann_isocam_2000, peretto_initial_2009}, where we do not measure the extinction of individual stars, but instead rely on measuring the attenuation against a diffuse and smoothly varying background light. Fundamentally, the technique measures the attenuation of the local stellar background caused by dust on a pixel-by-pixel basis. In order to apply this technique to other galaxies, we use readily available, high resolution optical \textit{Hubble Space Telescope} (HST) data to reconstruct the stellar background, and determine the column of dust (and conversely gas) by comparing the observed intensity of each pixel against what the intensity would be if there were no dust to cause any extinction (i.e. the reconstructed stellar distribution). We also use \textit{Herschel Space Observatory} \citep{pilbratt_herschel_2010} observations of dust emission to calibrate the unknown contribution of foreground light in the line-of-sight. This novel technique allows us to map entire galaxies at sub-arcsecond resolution, practically an order of magnitude better than the typical resolution achieved by current interferometric observations of entire nearby galaxies (generally a few arcseconds).

In this paper, we showcase our new high-resolution extinction mapping technique for our test galaxy M51 (NGC 5194), chosen for its ample survey coverage, its relatively near distance, and its near face-on inclination. We adopt 7.6 Mpc as the M51 distance to the MW \citep{ciardullo_planetary_2002}, and an inclination of $22^{\circ}$ \citep{tully_kinematics_1974, colombo_pdbi_2014_moment}. The data used are described in §\ref{sec:data}. In §\ref{sec:method} we describe the details of our high-resolution extinction technique. We present the resulting sub-arcsecond gas mass surface density map in §\ref{sec:hr_sd}, as well as a comparison with other dust mass and optical depth determinations in M51, and also with $^{12}$CO\,(1-0) observations from the PdBI Arcsecond Whirpool Survey \citep[PAWS;][]{schinnerer_pdbi_2013}. We provide a summary of our findings in §\ref{sec:sum_conc}.

\begin{figure}
    \centering
    \includegraphics[width=0.5\textwidth]{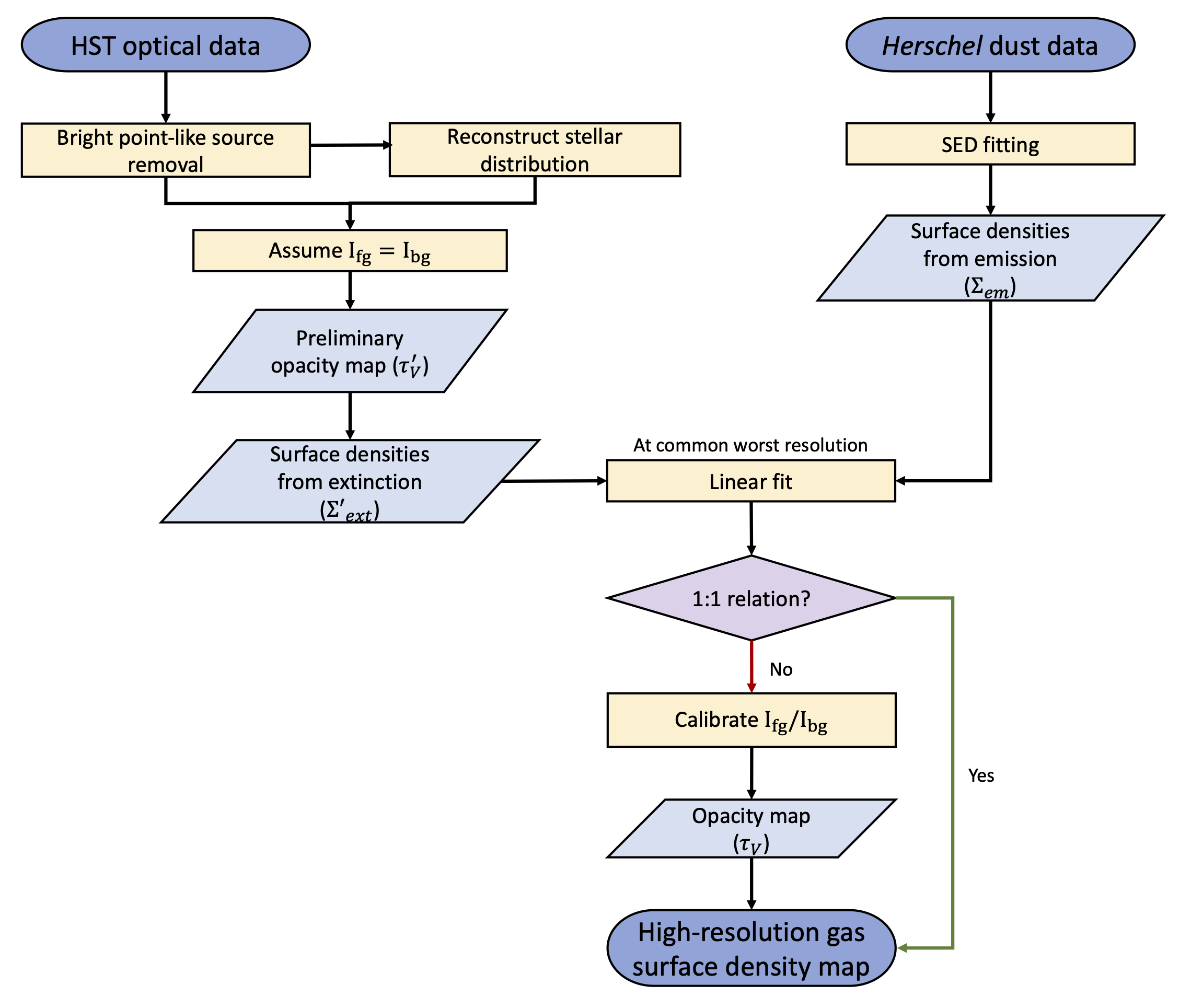}
    \caption{Simplified flowchart of the high-resolution extinction mapping technique presented in this paper (further details in §\ref{sec:method}).}
    \label{fig:flowchart}
\end{figure}

\section{Data}
\label{sec:data} % used for referring to this section from elsewhere

In this paper, we apply our new high-resolution extinction mapping method to HST optical data of M51. This technique measures dust extinction from HST optical data in order to build gas mass surface density maps. We use \textit{Herschel} dust emission observations as a calibrator for the unknown foreground light contribution along the line-of-sight, and CO data from PAWS to compare our results.

\subsection{Archival HST data}
\label{sec:hst}

The F435W (B-band, with pivot wavelength $433$ nm) and the F555W (V-band, with pivot wavelength $536$ nm) filters from the HST ACS (Advanced Camera for Surveys) Heritage, retrieved from the Hubble Legacy Archive\footnote{\url{https://hla.stsci.edu/}} are used here to build extinction maps of M51 in the optical. Both the B- and the V-band have a pixel size of $0.049"$ and resolution $0.1"$ \citep{mutchler_hubble_2005}. An astrometric correction of $0.1"$ in the RA direction and $-0.4"$ in the Dec. direction is applied to both the B-band and the V-band, as explained in \cite{schinnerer_pdbi_2013} - see also \cite{mutchler_hubble_2005}.

The V-band is chosen as the prime wavelength for our high-resolution extinction mapping technique as it should reach higher $A_V$ (visual extinction) before reaching saturation compared to the B-band, therefore probing deeper into the structure of the clouds. The B-band is used for testing the applicability of the method on a different band.

For the purpose of this paper, as a proof of concept, we are only interested in applying our new technique to NGC 5194 (referred to here as simply M51). Its companion, NGC 5195, is therefore cut from the original HST observations since it is known to have different metallicity and dust-to-gas mass ratio than M51 \citep[e.g.][]{mentuch_cooper_spatially_2012}. Additionally, we mask out the outer edges of the HST M51 observations using the reprojected \textit{Herschel} data as a guide.

\subsection{Archival \textit{Herschel} data}
\label{sec:herschel}

The high-resolution extinction mapping technique presented here allows us to map surface densities for nearby galaxies at an angular resolution close to the native resolution of HST. However, since we cannot make the simple assumption that all observed light is being attenuated by dust, we make use of \textit{Herschel} dust emission observations to retrieve the corresponding total dust column densities at a lower resolution and calibrate the contribution of foreground light  along the line-of-sight that we should assume for our technique so that statistically, the column densities obtained via dust emission and dust extinction measurements are consistent with each other.

At the IR-submm wavelengths where dust emission is the brightest, the spectral energy distribution (SED) of the galaxy is often modelled using a single-temperature modified blackbody (MBB) fit. In applying a single MBB fit, we assume: (1) the dust emission is optically thin, (2) the galaxy's population of dust grains is homogeneous in size and composition, (3) the dust emission seen comes primarily from heated large dust grains that are at thermal equilibrium with the interstellar radiation field and thus share an average temperature along the line of sight \citep[e.g.][]{draine_interstellar_2003,galliano_nearby_2022}, and (4) the mass absorption coefficient, $\kappa(\lambda)$, is a power-law in the shape $\kappa(\lambda) = \kappa_0 \ (\lambda_0 / \lambda)^\beta$ \citep{hildebrand_determination_1983}, with $\kappa_0$ being the value of $\kappa(\lambda)$ at some reference wavelength $\lambda_0$, and $\beta$ being the dust emissivity spectral index. Despite all these assumptions, single-temperature MBB fits are commonly used, and can reliably retrieve estimates of dust properties for $\lambda \geq 100 \, \upmu$m \citep{bianchi_vindicating_2013}.

In this paper, we perform a single-temperature, MBB SED fit to four bands from \textit{Herschel}, retrieved from the \textit{Herschel} Science Archive\footnote{\url{http://archives.esac.esa.int/hsa/whsa/}}. The bands used are the $160 \, \upmu$m band from the PACS instrument \citep[Photodetector Array Camera and Spectrometer for \textit{Herschel},][]{poglitsch_photodetector_2010}, and the $250 \, \upmu$m, $350 \, \upmu$m, and $500 \, \upmu$m bands from SPIRE \citep[Spectral and Photometric Imaging Receiver,][]{griffin_herschel-spire_2010}. The archival PACS and SPIRE data used here are the level 2.5 and level 3 data products, respectively \citep{davies_dustpedia_2017}. The pixel sizes for the $160 \, \upmu$m, $250 \, \upmu$m, $350 \, \upmu$m, and $500 \, \upmu$m bands are, respectively, $3.2"$, $6"$, $10"$ and $14"$, with the corresponding full width at half-maximum (FWHM) of $\sim11.5"$, $\sim18"$, $\sim25"$ and $\sim36"$. All SPIRE bands were converted from MJy/sr to Jy/pixel. The PACS $160\upmu$m data was chosen as the template pixel scale - remaining bands were therefore reprojected onto a $3.2"$/pix grid. All bands were convolved to the angular resolution of the SPIRE $500\upmu$m band (worst resolution), resulting in all four dust emission maps used here having a common FWHM of $36"$. The Gaussian kernels used in this operation were $34"$ (for the $160 \, \upmu$m), $31"$ (for the $250 \, \upmu$m) and $26"$ (for the $350 \, \upmu$m).    

The pixel-by-pixel SED fitting used non-linear least squares to estimate the dust temperatures ($T_d$) and dust emission surface densities ($\Sigma_{\text{em}}$) in each pixel of M51. The reference absorption coefficient adopted was $\kappa_{250\upmu\text{m}}=21.6$ cm$^2$g$^{-1}$ from \citet[hereafter OH94]{ossenkopf_dust_1994}, often used in MW studies \citep[e.g.][]{kauffmann_mambo_2008,2009_schuller_atlasgal}, which is valid for grains covered by a thick ice mantle that follow the classical Mathis, Rumpl, and Nordsieck (MRN) size distribution model \citep{mathis_size_1977}. We assume a fixed dust emissivity spectral index of $\beta=2$ - an appropriate assumption for metal-rich, late-type galaxies such as M51 \citep[see][]{boselli_far-infrared_2012,mentuch_cooper_spatially_2012, clark_empirical_2016}. The average values resulting from this SED fit are $T_d\simeq21.7\pm3.64$\,K and $\Sigma_{\text{em}}\simeq 0.06 \pm 0.03 \, \sunpc$. For comparison purposes, the dust surface densities obtained using $\kappa_{250\mu m}=3.98$\,cm$^{2}$g$^{-1}$ from \cite{mentuch_cooper_spatially_2012} are about a factor 5.4 larger than ours. Similarly, using $\kappa_{250\mu m}=10$\,cm$^{2}$g$^{-1}$ from \cite{elia_first_2013} (a Galactic study), the resulting dust surface densities are about a factor 2.2 larger.

\subsection{Archival PAWS data}
\label{sec:paws}

In this paper, we also compare our extinction results to the column densities derived from the $^{12}$CO\,(1-0) emission maps from PAWS \citep{schinnerer_pdbi_2013}, retrieved from the PAWS website\footnote{\url{https://www2.mpia-hd.mpg.de/PAWS/}; see also \url{https://www.iram-institute.org/EN/content-page-240-7-158-240-0-0.html}}. PAWS imaged the CO gas in the central 11 x 7 kpc region of M51 using both the Plateau de Bure Interferometer (PdBI) and the 30m single dish telescope from IRAM. The resulting combined datacubes (PdBI+30m) have a resolution of $1.16" \times 0.97"$ (corresponding spatial resolution of $\sim 40$ pc), a pixel scale of $0.3"$/pix, and a sensitivity of 0.4 K per 5 km s$^{-1}$ channel. 

\begin{figure*}
    \centering
    \includegraphics[width=\textwidth]{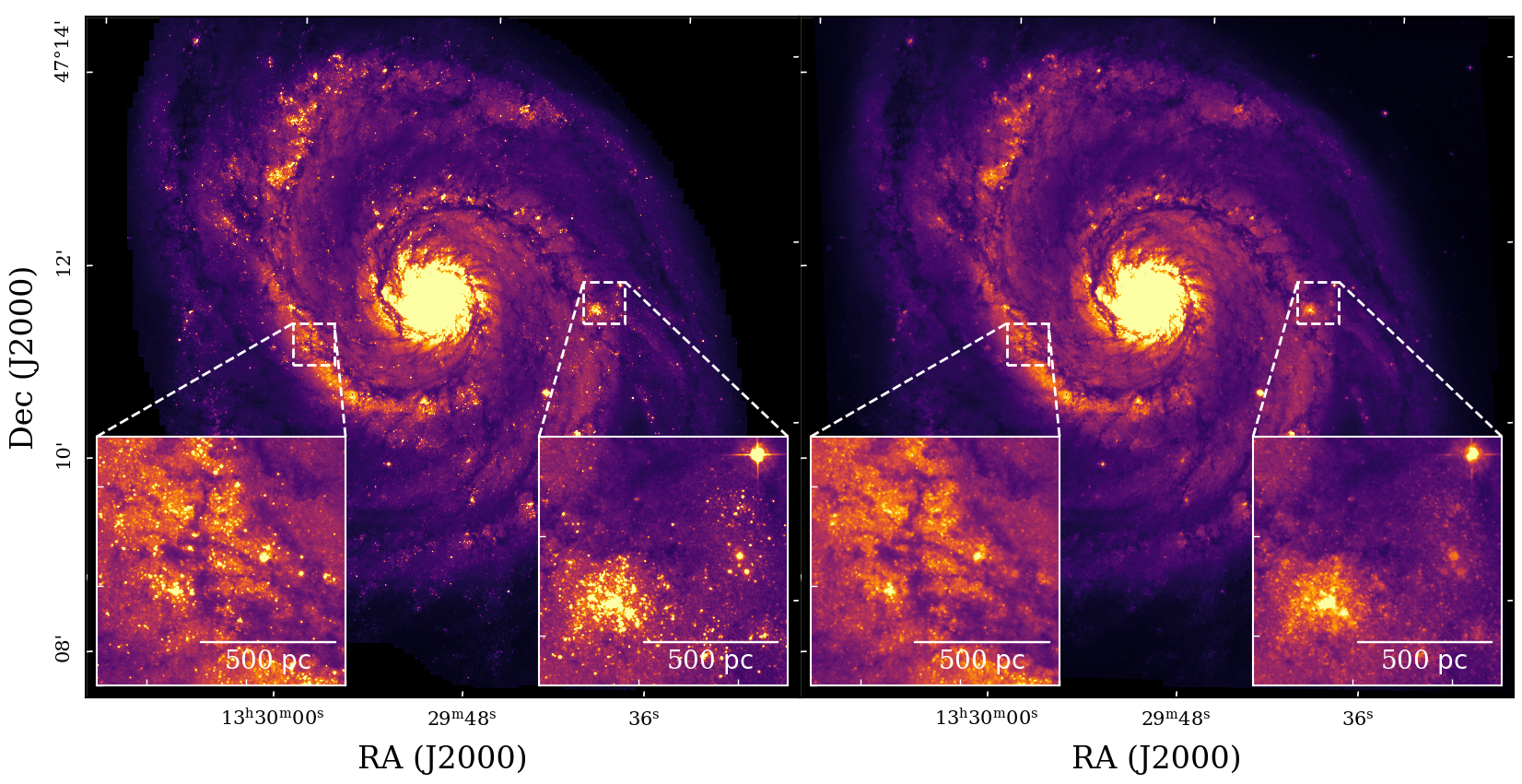}
    \caption{\textit{Left:} original HST V-band image of M51. \textit{Right:} M51 after removal of bright point-like sources. At the bottom of both panels are the same two regions zoomed for better visualisation of the source-removal results. Given that star clusters are not point-like sources, they are not removed from our image completely (as seen in the right panel). Additionally, in regions immediately on top of bright, young star clusters, we do not expect to retrieve molecular clouds.}
    \label{fig:source_removal}
\end{figure*}

\section{High-resolution extinction technique}
\label{sec:method}

We present a new high-resolution extinction mapping technique using HST data at visible wavelengths that allows us to construct an all-galaxy dust surface density map at a resolution close to the native resolution of HST. In summary, this technique is similar to the techniques used in Galactic studies in the IR \citep[e.g.][]{bacmann_isocam_2000,peretto_initial_2009}, which determine the amount of the local background radiation attenuated by the dust at each pixel. As such, this method only works if we can recover a reliable estimate of the local background radiation. This is only possible if the background radiation is smoothly varying - such that it can be reconstructed directly from the HST images using median filtering techniques. This is the case for the relatively diffuse stellar background traced by the B- and V-bands of HST, but it becomes less usable in the near IR bands due to the increased detection of younger, embedded stars, which dominate in the dust regions we are studying here, consequently causing the stellar background to be less smooth.

The different steps involved in this technique (summarised in Fig.~\ref{fig:flowchart}) are described in more detail in this section, namely the preparation of the images for the analysis (removing point-like sources), the reconstruction of the stellar background, computing the optical depth, and calibrating our results with lower-resolution surface density maps of our test galaxy, M51, obtained from dust emission observations from \textit{Herschel} at far-IR wavelengths.

\subsection{Removal of point-like sources}
\label{sec:removal}

The first step in our method consists of removing bright point-like sources, which can be the cause of artifacts in the final map. In particular, given that our method estimates the attenuation caused by the dust against a smoothly varying background on a pixel-by-pixel basis, in regions where bright point-like sources appear in the foreground, any dust attenuation that might exist behind is completely hidden by the bright object, effectively creating artificial "holes" of negative opacities in the background clouds. If these foreground stars are small enough (i.e. point sources of a size similar to the resolution of the HST image), we can remove them from the map and refill those positions with an interpolated background, which will give us an estimate of the opacity of the clouds behind the bright sources. Additionally, a high density of bright point-like sources (such as a large cluster of stars) can skew the median filtering techniques (see §\ref{sec:stellardist}), resulting in an overestimation of the average stellar distribution around those bright regions.

The simplest way to remove point sources from the original intensity would be to perform sigma-clipping and reject any pixels that are above a certain threshold. However, due to the wide dynamical range of intensities seen in the original HST map, this process is not straightforward. For example, if we use a relatively high value for the sigma - or if our median is skewed towards high values due to the really bright regions of the spiral arms or galactic centre - the sigma-clipping method will miss the fainter point-like sources clearly visible in the lower intensity regions such as the inter-arms. Conversely, if we use a lower value for the sigma-clipping, or the median, entire bright regions (like the galactic centre) are clipped off. To circumvent this, we need to perform the sigma-clipping on a map where background fluctuations are flattened, so that the sigma-clipping performs well across the entire map.

We use the \texttt{Background2D} function from \texttt{Photutils} \citep[v1.0.1,][]{larry_bradley_2020} to compute a flattened background to subtract from the original HST image in order to facilitate our source removal. Essentially, the function splits the original map into a grid and estimates the background of each square within the grid, ignoring all pixels that deviate more than 3 standard deviations ($\sigma$) in this calculation. The resulting grid is then transformed into the final, full-size estimated median background through bicubic spline interpolation. To capture the large background fluctuations we define a grid of $7.35"\times7.35"$ boxes ($\sim270\times270$ pc), which we transform into a full-size large-scale background map using a median filter with kernel size $0.245"$. Subtracting this large-scale background image from the original HST map does flatten the larger fluctuations of the original image, but many of the smaller variations are still included. Therefore, in an attempt to flatten our background as much as possible, we also estimate a small-scale background. This is done in similar manner but with a smaller grid size of $0.15"\times0.15"$ ($\sim5.50\times5.50$ pc).

Removing both large and small-scale backgrounds from the original HST V-band intensity leaves us with a map that includes only bright point-like sources surrounded by a relatively constant value. It is now possible to easily identify these point sources by creating a source mask using \texttt{make\_source\_mask} also from \texttt{Photutils}. This source mask contains only sources that have at least three connected pixels that are all $3\sigma$ above the median of the map. In practice, this procedure eliminates not only single point-like sources, but also clusters of stars, as well as any saturated HST source with diffraction spikes. Additionally, the bright pixels in the source mask are dilated with a $3 \times 3$ pixel square array to ensure that the whole bright source is clipped off rather than just the $> 3\sigma$ peaks.

Instead of simply removing the bright point-like sources in our final source mask, we fill it with an interpolated background value, which we obtain by convolving the HST V-band image (ignoring the bright point-like sources) with a Gaussian kernel of $\sim0.3"$, which is roughly equal to the size of the empty regions. We apply one final small convolution to account for any boundaries that might have been created in the filling process or any other artefacts. This convolution has a kernel size of $\sim0.1"$ and it only slightly degrades the resolution of the source-removed image to $0.14"$ ($\sim$5 pc at the distance of M51), compared to the $0.1"$ native angular resolution of the HST data. The original HST V-band image and the respective source-removed image are shown in Fig.~\ref{fig:source_removal}.

\subsection{Reconstruction of stellar distribution}
\label{sec:stellardist}

\begin{figure}
    \centering
    \includegraphics[width=0.4\textwidth]{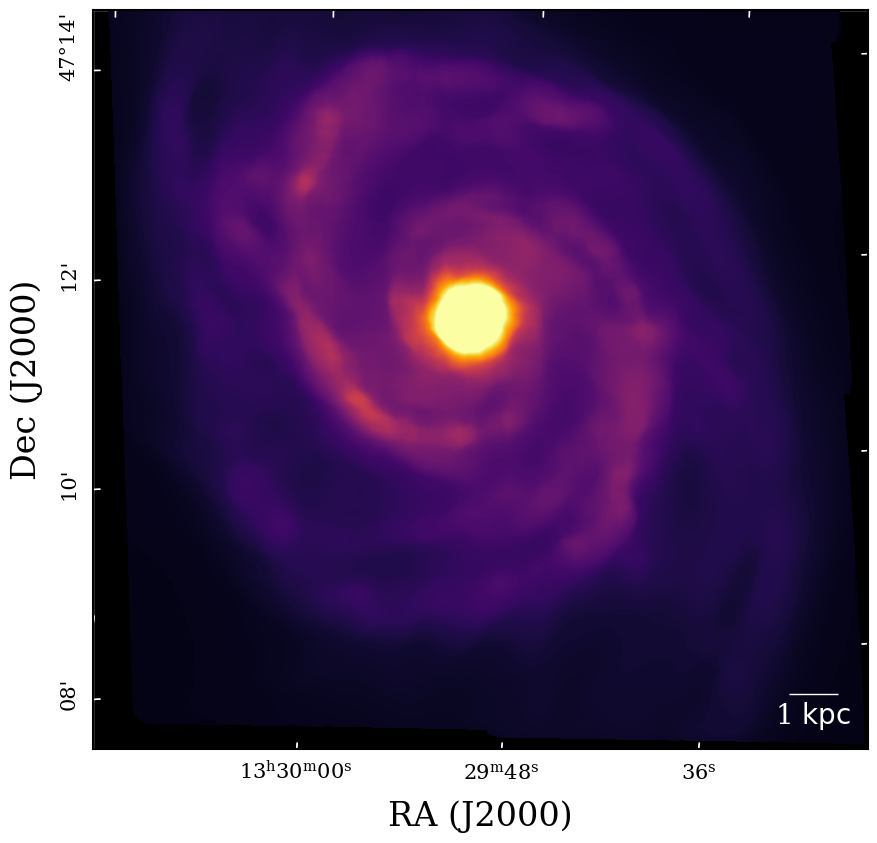}
    \caption{Reconstructed stellar distribution of M51 for the V-band. Extinction features present in the original HST V-band image are eliminated, whilst the overall shape and features (such as the bright spiral arms) of the galaxy are preserved.}
    \label{fig:stellar_distribution}
\end{figure}

The next step of our method is the reconstruction of the total stellar distribution, which is meant to reproduce the stellar radiation field of the galaxy if there was no light being absorbed by the dust grains in the ISM. The stellar distribution, $I_0$, is reconstructed by applying a sizeable median filter to the source-removed image, $I_\lambda$ (§\ref{sec:removal}). In order to have a good approximation of the total stellar light, the kernel must be bigger than the typical size of the absorbing structures (i.e. molecular clouds), so that the stellar light estimate is not sensitive to the absorbing medium, but not too big as to risk losing the intrinsic fluctuations of the stellar distribution due, for instance, to the spiral structure of the galaxy. The choice of the kernel is therefore crucial so as to not under- or over-estimate the stellar distribution, as that will result in under- and over-estimating the respective opacities. The kernel size chosen was $\sim16"$ ($\sim$600pc) - a little under the size of the galactic nuclear bar for M51 \citep[see][]{colombo_pdbi_2014}. This scale is larger than the typical size of giant molecular clouds, but small enough to capture the changes due to spiral arms and the galactic centre. Changing this kernel by a factor 2 either way - 300pc or 1.2kpc - changes the final calibrated opacities by a factor 1.1 and 0.97, respectively. The final reconstructed $I_0$\footnote{Assuming a fixed total stellar mass for M51, ${M_{*} = 4.5\times10^{10} \, \text{M}_\odot}$ \citep{leroy_z_2019}, it is possible to retrieve an effective mass-to-light ratio and thus convert $I_0$ into a stellar mass surface density estimate. Our total stellar mass estimate is a factor of 1.26 and 1.23 higher than the values quoted by \cite{querejeta_spitzer_2015} and \cite{martinez-garcia_removing_2017}, respectively, well within the expected uncertainties.} is shown in Fig \ref{fig:stellar_distribution}.

\subsection{Constructing a preliminary opacity map}
\label{sec:opacity}

Dust absorbs radiation in the optical. The simplest approach to describe how the observed light, $I_\lambda$, is being attenuated by dust is by assuming a flat foreground screen of dust, such that: ${I_\lambda = I_{0} \ e^{-\tau_\lambda}}$, where $\tau_\lambda$ is the optical depth of the dust. In reality, dust is likely to be evenly mixed with the gas within a galaxy \citep[see][and references therein]{calzetti_dust_2001}, and the use of a foreground dust screen often overestimates the total attenuation \citep[e.g.][]{calzetti_dust_2001, kessler_pa_2020}. Here, we adopt a "sandwich" geometry model where the dust sits in a layer close to the mid-plane of the galaxy (which is a fair assumption for face-on galaxies, such as M51). This dust layer follows the distribution of the stellar light of the galaxy, but it does not include any clumps \citep[e.g.][]{tuffs_modelling_2004}. Considering this "sandwich" dust/stars geometry, the observed light can be described through:

\begin{equation}
    \centering
    \label{eqn:dustextinction}
    I_\lambda = I_{\text{bg}} \ e^{-\tau_\lambda} + I_{\text{fg}},
\end{equation}

\noindent where $I_\lambda$ (the point-like source removed-image, see §\ref{sec:removal}) is the sum of the light that is free to travel in the line-of-sight without any obscuration (i.e. foreground intensity, $I_{\text{fg}}$) and the light that has to travel through an absorbing medium to reach us (i.e. background intensity, $I_{\text{bg}}$). The stellar distribution, $I_0$, can be related to the foreground and background intensities through $I_0 = I_{\text{fg}} + I_{\text{bg}}$. Additionally: $I_{\text{bg}} = b\,I_0$, and $I_{\text{fg}} = f\,I_0$, where $b$ and $f$ are the background and foreground fractions of the total light respectively, and $b + f = 1$. Eq. (\ref{eqn:dustextinction}) can be rearranged to give the spatial distribution of the optical depth or line-of-sight opacity:

\begin{equation}
    \centering
    \label{eqn:method}
    \tau_\lambda = - \text{ln} \left( \frac{I_\lambda - I_{\text{fg}}}{I_{\text{bg}}} \right).
\end{equation}

\begin{figure}
    \centering
    \includegraphics[width=0.45\textwidth]{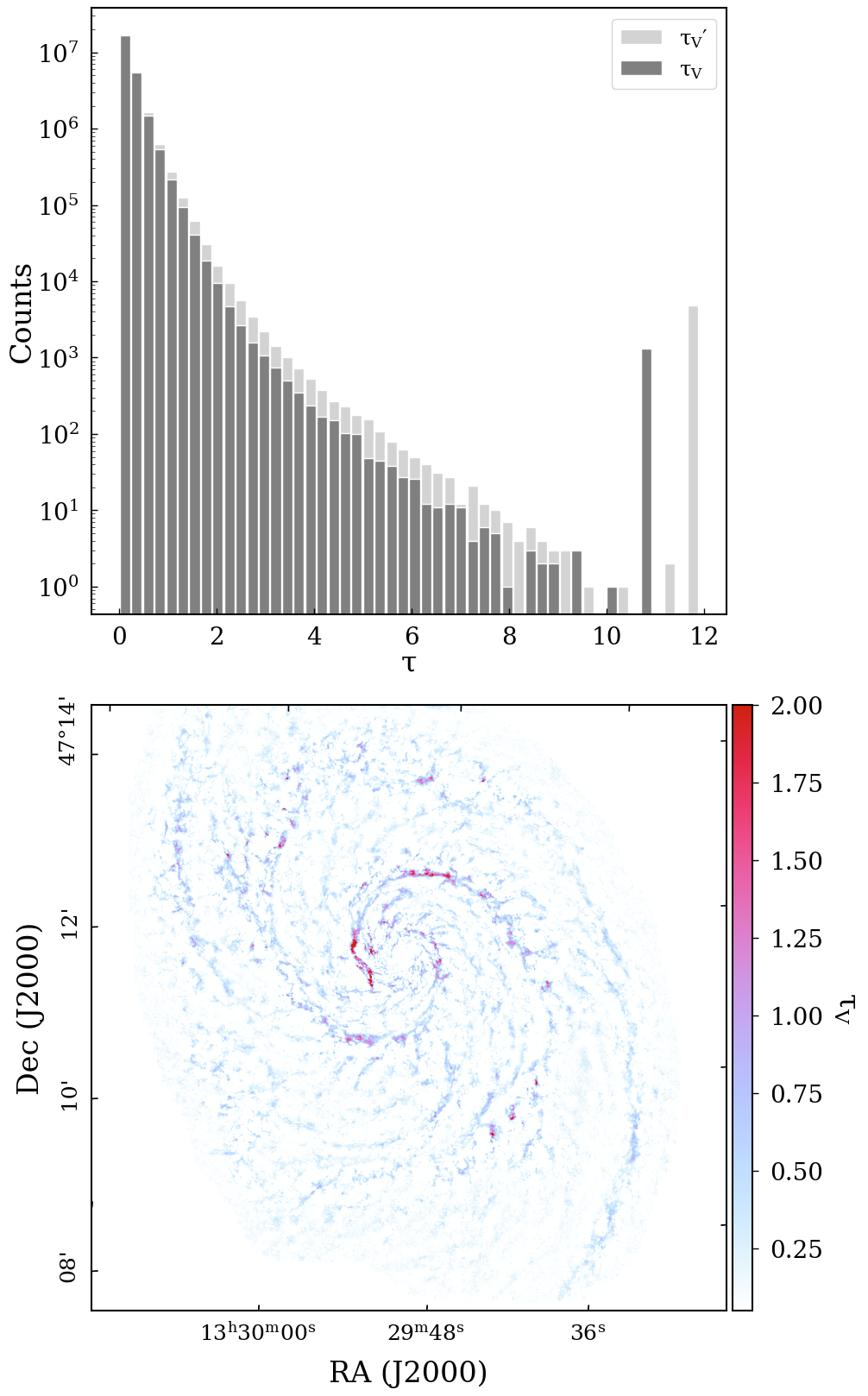}
    \caption{\textit{Top:} Histograms of first-guess optical depth, $\tau'_V$, (in light grey) and calibrated optical depth $\tau_V$, (in dark grey). The isolated bins at the higher end of each data set correspond to the saturated pixels of each map, set at 1\% higher than the maximum optical depth value. \textit{Bottom:} Spatial distribution of the optical depth of M51 resulting from the calibration process within our method (further details in §\ref{sec:calibration}). The calibrated optical depth, $\tau_V$, is built with $b=0.53$ and $f=0.47$ as the background and foreground light fractions, respectively.}
    \label{fig:preopacity}
\end{figure}

To construct a preliminary opacity map for the V-band, $\tau'_V$, it was assumed that $I'_{\text{bg}} = I'_{\text{fg}}$, and therefore $b'=f'=0.5$. We adopt a prime superscript on the quantities referring to our first guess of background/foreground fractions. Physically, this would mean that the emitting light is split equally beneath and above the absorbing dust content of the galaxy. We do not expect the background/foreground fraction to stray far from these values in a face-on spiral like M51, given that the gas scale height of a disc galaxy is typically smaller than that of the stars \citep[e.g.][]{patra_molecular_2019}, and therefore we expect most of the obscuring dust to lie close to the galactic plane. Still, since clouds become optically thick relatively quickly in the optical it would be possible to reach optical depths slightly offset from the exact mid-plane of the galaxy. It is also possible that the reconstructed stellar distribution is not an accurate representation of the true stellar distribution against which the clouds are absorbing (particularly where there are large patches of obscuring dust lanes that will skew the large median filters and underestimate the total stellar light), and therefore the background/foreground fractions might not lie exactly at 50\%. Nonetheless, this initial background/foreground fraction assumption is not very important, as it will be calibrated at a later stage (described in §\ref{sec:calibration}). The distribution of our first-guess opacity map, $\tau'_V$, is represented in Fig.~\ref{fig:preopacity} (top panel).

\subsection{Constraining $I_{\text{fg}}$ and $I_{\text{bg}}$}
\label{sec:calibration}

It is impossible to constrain $I_{\text{bg}}$ and $I_{\text{fg}}$ with the optical data alone. However, dust emission observations in the IR-submm wavelength range can provide an independent measurement of the dust mass surface densities, albeit at lower resolution. Thus, we perform a comparison between our dust extinction maps and dust emission surface densities fitted from \textit{Herschel} observations (§\ref{sec:herschel}) in order to calibrate our technique and automatically constrain the background/foreground fraction.

Before proceeding with this comparison, we first need to deal with the regions where our $\tau'_V$ estimates are not usable: negative optical depth pixels and ``saturated'' pixels. Pixels with negative $\tau'_V$ were masked to zero, since these would correspond to points where the observed intensity is brighter than the assumed background stellar distribution and therefore there is no measurable dust attenuation. In addition, in pixels where the observed intensity is lower than the assumed foreground emission, we obtain undefined values (NaNs). We call these our "saturated" regions, and fill them with a fixed maximum value set at $1\%$ higher than the peak value of $\tau'_V$. Less than $0.01 \%$ of the pixels in our opacity map are saturated.

The optical depth can be related to dust surface densities, $\Sigma_{\text{dust}}$, through a dust mass absorption coefficient at the relevant wavelength, $\kappa_\lambda$, following:
\begin{equation}
    \centering
    \label{eqn:opkappasd}
    \tau_\lambda = \kappa_\lambda \, \Sigma_{\text{dust}}.
\end{equation}

\noindent Since we are working with dust extinction in the V-band, the above equation becomes $\tau'_V = \kappa_{V} \, \Sigma'_{\text{ext}}$, where $\tau'_V$ is our preliminary opacity, $\kappa_V$ is our adopted dust absorption coefficient \citep[$\kappa_V=8.55\times10^3$ cm$^2$ g$^{-1}$ , or equivalently $\kappa_V=1.79$ pc$^2$ M$_\odot^{-1}$,][hereafter D03]{draine_interstellar_2003}, and $\Sigma'_{\text{ext}}$ is our preliminary extinction surface densities.

\begin{figure*}
    \centering
    \includegraphics[width=0.8\textwidth]{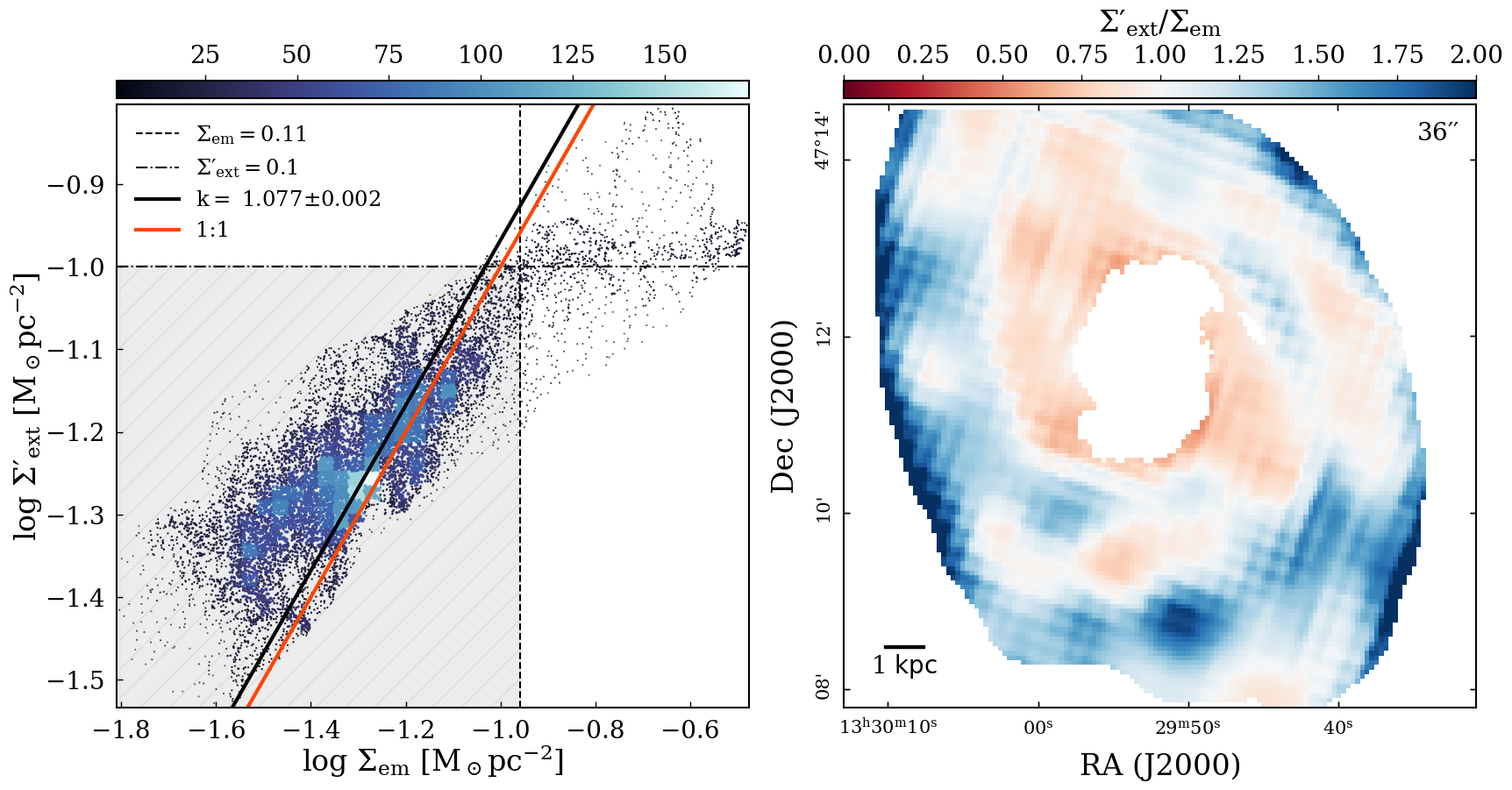}
    \caption{\textit{Left}: First-guess dust surface densities from our extinction technique, $\Sigma'_{\text{ext}}$, against the surface densities estimated from \textit{Herschel} dust emission data, $\Sigma_{\text{em}}$. Both quantities are in logarithmic space. The blue scale represents the density of points. The significant scatter seen in the plot corresponds to regions where dust extinction and dust emission are not sensitive to the same column. We perform a linear fit where this scatter is minimal (grey shaded region in plot), shown as a solid black line. The desired 1:1 relation is plotted as a solid red line. \textit{Right}: Ratio of the preliminary extinction dust surface densities, $\Sigma'_{\text{ext}}$, and the surface densities fitted from dust emission \textit{Herschel} observations, $\Sigma_{\text{em}}$, done at the resolution and pixel scale of $\Sigma_{\text{em}}$ ($36"$ and $3.2"$/pix, respectively). Only the points below $\Sigma'_{\text{ext}} = 0.1 \, \sunpc$ and $\Sigma_{\text{em}} = 0.11 \, \sunpc$ (i.e. grey-shaded region in left panel) are plotted, showing that the excluded points are exclusively in the centre of the galaxy.}
    \label{fig:extem_ratio}
    \label{fig:linear_fit}
\end{figure*}

The comparison between extinction and emission is done in the grid ($3.2"$/pix) and resolution ($36"$) of the coarser dust emission column density map. Consequently, the $\Sigma'_{\text{ext}}$ is regridded and convolved accordingly.

The left panel of Fig.~\ref{fig:linear_fit} shows the relationship between the convolved, preliminary extinction surface densities from our method, $\Sigma'_{\text{ext}}$, and the emission surface densities from \textit{Herschel}, $\Sigma_{\text{em}}$ (both at $36"$ resolution). It is clear from the figure that the relationship between the dust extinction and emission surface densities is not exactly linear, with a systematic flattening of $\Sigma'_{\text{ext}}$ towards higher values of surface density. This is expected, as estimates of column density from dust extinction and emission do not always match exactly. In the densest regions with high surface densities, the dust becomes optically thick quickly at visible wavelengths, and so the surface densities derived from dust extinction will be a lower estimate of the column, while dust emission should still be sensitive to the full column. On the other hand, for the more diffuse regions with lower surface densities and for regions where the dust is being heated (such as the centre of M51, see Appendix \ref{sec:appC}), the ISM is warmer and thus the SED will start peaking at shorter wavelengths (and outside the wavelengths used for our SED fitting), causing the surface densities from the dust emission to be less accurate. In terms of spatial agreement between dust emission and dust attenuation, \cite{thilker_phangsjwst_2023} recently analysed the spatial distribution of attenuation from the HST B-band against several wavelengths tracing dust emission from PHANGS-JWST\footnote{\url{http://www.phangs.org}} \citep{lee_phangsjwst_2023}. They found that in NGC 628, over $40\%$ of sight lines contain both dust emission and extinction at 25 pc scales (increasing to $55\%$ at 200 pc). Furthermore, they argue that this may be a lower estimate due to line-of-sight effects and difficulties in consistently extracting filamentary structures. In this work, we are not able to retrieve any measurement of dust extinction wherever star clusters sit in front of the gas/dust, whereas dust emission can measure the full column. Still, we perform our comparison between extinction and emission at a spatial resolution of about 1 kpc, where any significant smaller scale variations should average out, and thus differences in column sensitivity between the two tracers should be predominant. These sensitivity differences are in fact what is seen in the right panel of figure \ref{fig:extem_ratio}, which shows the spatial distribution of the ratio between the two surface density maps, $\Sigma'_{\text{ext}} / \Sigma_{\text{em}}$. As expected, in regions of high density (i.e. spiral arms, shown in red), $\Sigma'_{\text{ext}} < \Sigma_{\text{em}}$, and in more diffuse regions (shown in blue) $\Sigma'_{\text{ext}} > \Sigma_{\text{em}}$. 

In order to calibrate our extinction surface densities with the emission ones so that the final calibrated extinction map is (statistically) consistent with the emission map, we start by analysing the relation between our first-guess of $\Sigma'_{\text{ext}}$ and the reference $\Sigma_{\text{em}}$ from the emission. In this comparison we choose to ignore the centre of M51 due to not only potential temperature effects (further explained in Appendix \ref{sec:appC}) but also saturation effects. Considering only the points where $\Sigma'_{\text{ext}} < 0.1 \, \sunpc$ and $\Sigma_{\text{em}} < 0.11 \, \sunpc$ (shaded region shown in left panel of Fig.~\ref{fig:linear_fit}), a linear regression is a reasonable approach within the scatter. A simple linear fit gives a slope of $k = 1.077 \pm 0.002$. Changing the cutoff values by 30\% changes the slopes of the linear fit only marginally: $k=1.030\pm0.002$ for $\Sigma'_{\text{ext}} < 0.13 \, \sunpc$ and $\Sigma_{\text{em}} < 0.14 \, \sunpc$, and $k=1.14\pm0.003$ for $\Sigma'_{\text{ext}} < 0.07 \, \sunpc$ and $\Sigma_{\text{em}} < 0.08 \, \sunpc$. We use the $k = 1.077 \pm 0.002$ fit to make a new decision on the background/foreground fraction, and thus calibrate our optical depths.

Neglecting nonlinearity (i.e. where dust extinction and emission are not sensitive to the same column), our calibrated extinction surface densities, $\Sigma_{\text{ext}}$, should equal $\Sigma_{\text{em}}$. Knowing that our preliminary $\Sigma'_{\text{ext}}$ relate to the emission surface densities through $\Sigma'_{\text{ext}} = k \, \Sigma_{\text{em}}$, with $k$ being the aforementioned slope of the linear fit, then it is possible to write a relation between our first estimate and our calibrated estimate of extinction dust surface densities:
\begin{equation}
    \label{eqn:sdextsdem}
    \Sigma_{\text{ext}} = \frac{\Sigma'_{\text{ext}}}{k}.
\end{equation}
 
\noindent This relation (and consequently $k$) tells us by how much we should change our measured extinction surface densities such that they match the emission on a global scale (at lower resolution). We assume that this global "fractional" change remains the same for the finer grid of our high-resolution map, and use this to re-calibrate our $b$ and $f$ fractions. 
  
Combining Eq. (\ref{eqn:opkappasd}) with Eq. (\ref{eqn:sdextsdem}), gives $\tau_{V} = \tau_{V}' / k$. Replacing Eq. (\ref{eqn:method}) into the previous expression gives:
\begin{equation}
    \centering
    \label{eqn:step1}
    \frac{I_{V} - I_{\text{fg}}}{I_{\text{bg}}} = \left( \frac{I_{V} - I'_{\text{fg}}}{I'_{\text{bg}}} \right)^{1/k}, 
\end{equation}

\noindent which relates the new calibrated opacity ($\tau_{V}$) to the preliminary opacity ($\tau'_{V}$, see §\ref{sec:opacity}). Remembering that $I_{\text{bg}} = b \, I_0$, and likewise $I_{\text{fg}} = (1 - b) \, I_0$, it is possible to rearrange Eq. (\ref{eqn:step1}) to give the new calibrated background distribution, $b$:  

\begin{equation}
    \centering
    \label{eqn:b}
    b = \frac{I_V - I_0}{I_0 \, \left[ \left( \frac{I_V - I'_{\text{fg}}}{I'_{\text{bg}}} \right)^{1/k} - 1 \right]},
\end{equation}

\noindent where $I_V$ is the source-removed intensity for the V-band (§\ref{sec:removal}), $I_0$ is the reconstructed stellar distribution (§\ref{sec:stellardist}), $I'_{\text{fg}}$ and $I'_{\text{bg}}$ are our first-guess foreground and background intensities with $f' = b' = 0.5$ (§\ref{sec:opacity}), and $k$ is the slope of the linear fit between $\Sigma'_{\text{ext}}$ and $\Sigma_{\text{em}}$. Eq. (\ref{eqn:b}) gives, for each pixel in our high-resolution map, what the background fraction $b$ (and consequently the foreground fraction since $f = 1 - b$) should be so that the surface densities derived from our dust attenuation technique correlate with the kpc-scale surface densities computed from dust emission at IR to sub-mm wavelengths.

We choose to adopt a single value for the calibrated $b$ and $f$, rather than use a pixel-by-pixel correction that would guarantee our convolved extinction surface densities match the \textit{Herschel} dust emission densities exactly. As explained before, we do not expect dust extinction and emission to always be sensitive to the same column densities. Furthermore, in regions where dust emission is particularly bright (i.e. centre), dust temperatures and properties may be different than the rest of the galaxy, which will not be accounted for with our single-temperature and single-$\beta$ SED fits. In order to not propagate these ambiguities into the dust extinction - which is independent of dust temperature - we prefer to adopt a statistical approach to estimate the typical $b$ (and $f$) across the galaxy. This is further discussed in Appendix \ref{sec:appC}.

The medians of the calibrated background and foreground fraction are, respectively, $b=0.53$ and $f=0.47$, with a tight interquartile spread of $\text{Q}75-\text{Q}25 = 0.01$. Note that this background/foreground calibration process gives similar results for different initial assumptions of $b'$ and $f'$ (e.g. starting with $b' = 0.6$, the background fraction is calibrated to $b=0.51$, with $\text{Q}75-\text{Q}25 = 0.01$). 

To construct the new optical depth map for the V-band, $\tau_V$, we simply feed the calibrated $b$ and $f$ into Eq. (\ref{eqn:method}), keeping the remaining maps ($I_V$ and $I_0$) the same as before. The calibrated opacity map receives the same treatment described in §\ref{sec:opacity}: negative opacities are masked to zero, and saturated pixels are filled with a value corresponding to a $1\%$ increase of the peak value of $\tau_V$. The resulting $\tau_V$ distribution is shown in the bottom panel of Fig.~\ref{fig:preopacity}. We estimate that values of opacity within 30\% of our maximum value of $\tau$ (i.e. opacities close to saturation, $\tau > 7$) are more uncertain, and likely lower limits due to saturation effects, but a more detailed analysis is shown in Appendix \ref{sec:appA}.

\section{Final high-resolution gas surface density map}
\label{sec:hr_sd}

\begin{figure*}
    \centering
    \includegraphics[width=0.98\textwidth]{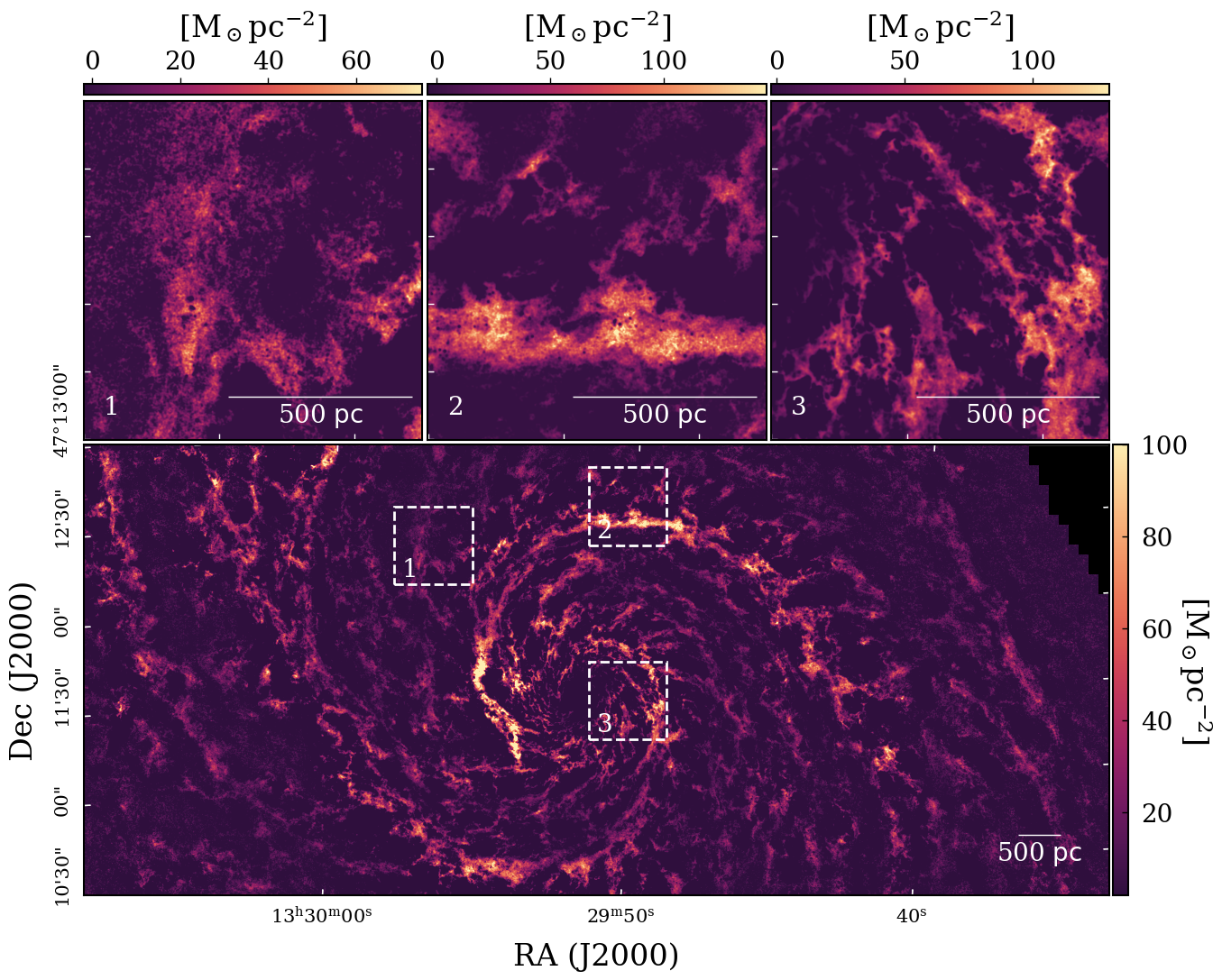}
    \caption{Gas surface density map of a portion of M51 resulting from our high-resolution extinction-mapping technique (\textit{bottom}), and with zoom-ins of three example regions in the inter-arms (1), spiral arm (2) and centre (3) shown in the top row.}
    \label{fig:sd}
\end{figure*}

Recalling Eq. (\ref{eqn:opkappasd}), it is possible to convert our calibrated opacities, $\tau_V$ (§\ref{sec:calibration}), into dust surface densities through a dust absorption coefficient, $\kappa_V=1.786$ pc$^2$ M$_\odot^{-1}$ (D03). Additionally, assuming a dust-to-gas mass fraction of 0.01, and that the gas and dust are well-mixed, the dust surface densities are easily converted into their gas equivalent. The resulting gas surface density map, $\Sigma_{\text{gas}}$, will inherit the resolution and pixel scale of the optical depth map ($0.14"$ and $0.049"$/pix, respectively), which translates into a spatial resolution of $\sim 5$ pc for M51 at the adopted distance of $7.6$ Mpc \citep{ciardullo_planetary_2002}. Fig.~\ref{fig:sd} showcases the final high-resolution gas surface density distribution resulting from our extinction mapping technique for M51\footnote{This map is publicly available at \url{https://dx.doi.org/10.11570/23.0010}, and also at the FFOGG (Following the Flow of Gas in Galaxies) project website (\url{https://ffogg.github.io/ffogg.html}).}. The uncertainty in the opacities mentioned in the previous section (§\ref{sec:calibration}) will be inherited by the surface densities; we perform Monte Carlo simulations to obtain relative uncertainties, as well as determine an upper limit of surface densities we can accurately measure, all of which is further detailed in Appendix \ref{sec:appA}.

\subsection{Impact of dust absorption coefficient assumption}
\label{sec:differentkappa}

As previously stated, our extinction technique employs the D03 specific opacity for the V-band and the OH94 specific opacity at intermediate densities for the IR. Consequently, our calibrated gas surface densities will reflect these dust estimates. In order to calibrate the background/foreground split of stellar light in relation to the dust (§\ref{sec:calibration}), our technique requires that the opacity laws used in the two different wavelength regimes (i.e. visual and IR) be consistent with each other. Keeping $\kappa_V$ from D03 and instead adopting $\kappa_{250\upmu\text{m}} = 3.98$ cm$^2$g$^{-1}$ from \cite{mentuch_cooper_spatially_2012} to construct $\Sigma_{\text{em}}$, for example, would lead our calibration procedure to output a background fraction of just $b = 0.13$ (i.e. only the bottom $13\%$ of the stellar light in M51 would be attenuated), and our calibrated opacities would peak at values of roughly 14.3 (visual extinction values up to 15.5). Additionally, over a quarter of our final map would be saturated, preventing us from estimating an opacity value in a significant portion of M51.

Alternatively, we could fix our background/foreground light fraction and instead attempt to retrieve a new dust absorption coefficient in the optical that would match the $\kappa_{250\upmu\text{m}}$ from \cite{mentuch_cooper_spatially_2012} (or any other $\kappa_{250\upmu\text{m}}$). Remembering Eq. (\ref{eqn:opkappasd}), the new specific opacity in the visual, $\kappa_V^{\text{new}}$, would simply be the ratio between $\tau_V$ (assuming a fixed $b=0.53$ and $f=0.47$) and the dust emission surface densities derived using $\kappa_{250\upmu\text{m}}$ from \cite{mentuch_cooper_spatially_2012}, $\Sigma_{\text{em}}^{\text{new}}$. This results in a median ${\kappa_V^{\text{new}} = 0.345 \, \text{pc}^2 \, \text{M}_\odot^{-1}}$, or ${\kappa_V^{\text{new}} = 1.65 \times 10^{3} \, \text{cm}^2 \, \text{g}^{-1}}$. Utilising this combination of absorption coefficients (i.e. $\kappa_{250\upmu\text{m}}$ from \citealt{mentuch_cooper_spatially_2012} and the derived $\kappa_V^{\text{new}}$) results in a final gas surface density map with values a factor $\sim 5$ larger than our original map, but otherwise retains the same structure (i.e. the maps are proportional to each other).

It is possible to write a scaling relationship between our calibrated extinction surface densities, $\Sigma_{\text{ext}}$ (using OH94 and D03), and other dust absorption coefficients for the IR:

\begin{equation}
    \Sigma_{\text{ext}}^{\text{new}} = \left( \frac{21.6 \, \text{cm}^{2}\text{g}^{-1}}{\kappa^{\text{new}}} \right) \, \Sigma_{\text{ext}},
    \label{eq:scaling}
\end{equation}

\noindent where $\Sigma_{\text{ext}}^{\text{new}}$ is the high-resolution, extinction-derived gas surface density map for M51 resulting from calibrating to a different opacity law in the IR ($\kappa^{\text{new}}$ at $250\,\upmu\text{m}$). We caution that although this allows for the use of different dust absorption coefficients (and therefore different dust models), this scaling relationship is only true if the background/foreground fraction ($b/f$) is fixed at $b = 0.53$ and $f = 0.47$. Since the relationship between $b/f$ and the assumed opacity laws is not linear, it is not possible to calibrate for both $b/f$ and $\kappa$ at the same time. In this paper, we prefer to adopt an established value of $\kappa$ from the literature and calibrate for $b/f$ instead.

As previously stated, adopting a different combination of specific opacity laws will only change our final gas surface density map by a scaling factor; the observed structural hierarchy, or environmental trends, will remain the same. As will be shown in the following sections, after correcting for the difference in dust recipes used (i.e. applying a scaling factor), our high-resolution extinction mapping technique is able to retrieve similar gas/dust surface density (and optical depth) values to estimates of other studies which apply various independent approaches.

\subsection{Comparison with other dust studies}
\label{sec:comparison_otherdust}

The dust extinction technique outlined in this paper is one of many ways of measuring the dust mass of galaxies \citep[see e.g.][]{calzetti_dust_2001}. One such method involves measuring the reddening of stellar light in order to retrieve colour-based extinction maps \citep[e.g.][]{regan_overluminous_2000, thompson_hubble_2004, kainulainen_determination_2007}. For M51, \cite{holwerda_gaps_2007} constructed a colour map using near- to mid-IR bands, \textit{(I - L)}, and found an average optical depth of $\tau_{(I-L)}=0.18 \pm 0.09$ within one of the areas imaged by WFPC2 aboard HST. For roughly the same region in our map (coverage differs due to our cropping of NGC 5195), we report a similar median optical depth ($\tau_V = 0.15 \pm 0.1$).)

Colour-based extinction approaches often assume a simple foreground dust screen geometry and a single colour (and consequently age) for the stellar population, with the resulting dust masses being systematically lower than estimates from attenuation of ionised gas emission lines \citep[e.g.][]{kreckel_mapping_2013}. An alternative method to measure the dust mass within a galaxy is to perform radiative transfer simulations. These high-resolution models can account for more complex dust/stars geometry, as well as different dust heating sources (e.g. young ionising stars, old stellar population, AGN). \cite{nersesian_high-resolution_2020} utilise a state-of-the-art 3D radiative transfer simulation, which includes the \texttt{THEMIS} dust model \citep{jones_evolution_2013, jones_global_2017}, and the \texttt{CIGALE} SED fitting routine \citep{noll_analysis_2009, boquien_cigale_2019}. They find that the M51 system holds a total dust mass of $M_{\text{dust}} = 3.40 \, (\pm 0.65) \times 10^7$ M$_\odot$ for a dust absorption coefficient of $\kappa_{250\upmu\text{m}} = 6.40\,\text{cm}^{2}\,\text{g}^{-1}$ (adapted from the \texttt{THEMIS} model). Our high-resolution extinction technique retrieves a dust mass for M51 (i.e. just NGC 5194) of $M_{\text{dust}} = 9.70 \times 10^6\,\text{M}_\odot$. As discussed above (§\ref{sec:differentkappa}) and shown in Eq. (\ref{eq:scaling}), we can apply a scaling factor to our estimates of dust surface density to account for a different dust absorption coefficient assumed for the IR. Using the specific opacity from the \texttt{THEMIS} model would scale our surface densities by roughly a factor 3.4, resulting in a final dust mass of $M_{\text{dust}} = 3.27 \times 10^7\,\text{M}_\odot$, which is consistent with the dust mass range found by \cite{nersesian_high-resolution_2020}\footnote{Although we caution that in the \texttt{THEMIS} model, a $\beta=1.79$ is assumed rather than $\beta=2.0$.}. Using a similar radiative transfer simulation but with the \cite{draine_infrared_2007} dust recipe, \cite{de_looze_high-resolution_2014} retrieve a dust mass of $M_{\text{dust}} = 7.70 \times 10^7\,\text{M}_\odot$ for M51. Again following Eq. (\ref{eq:scaling}), our estimate for the total dust mass would increase by roughly a factor 5 if we had adopted the \cite{draine_infrared_2007} absorption coefficient in our dust emission modelling (equivalent to $\kappa_{250\upmu\text{m}} = 3.98\,\text{cm}^2\,\text{g}^{-1}$, \citealt{mentuch_cooper_spatially_2012}), which is within a factor 0.7 of the mass value quoted by \cite{de_looze_high-resolution_2014}.

\subsection{Comparison with PAWS integrated line intensity, $W_{\text{CO}}$}

\label{sec:paws_comparison}

\begin{figure*}
    \centering
    \includegraphics[width=\textwidth]{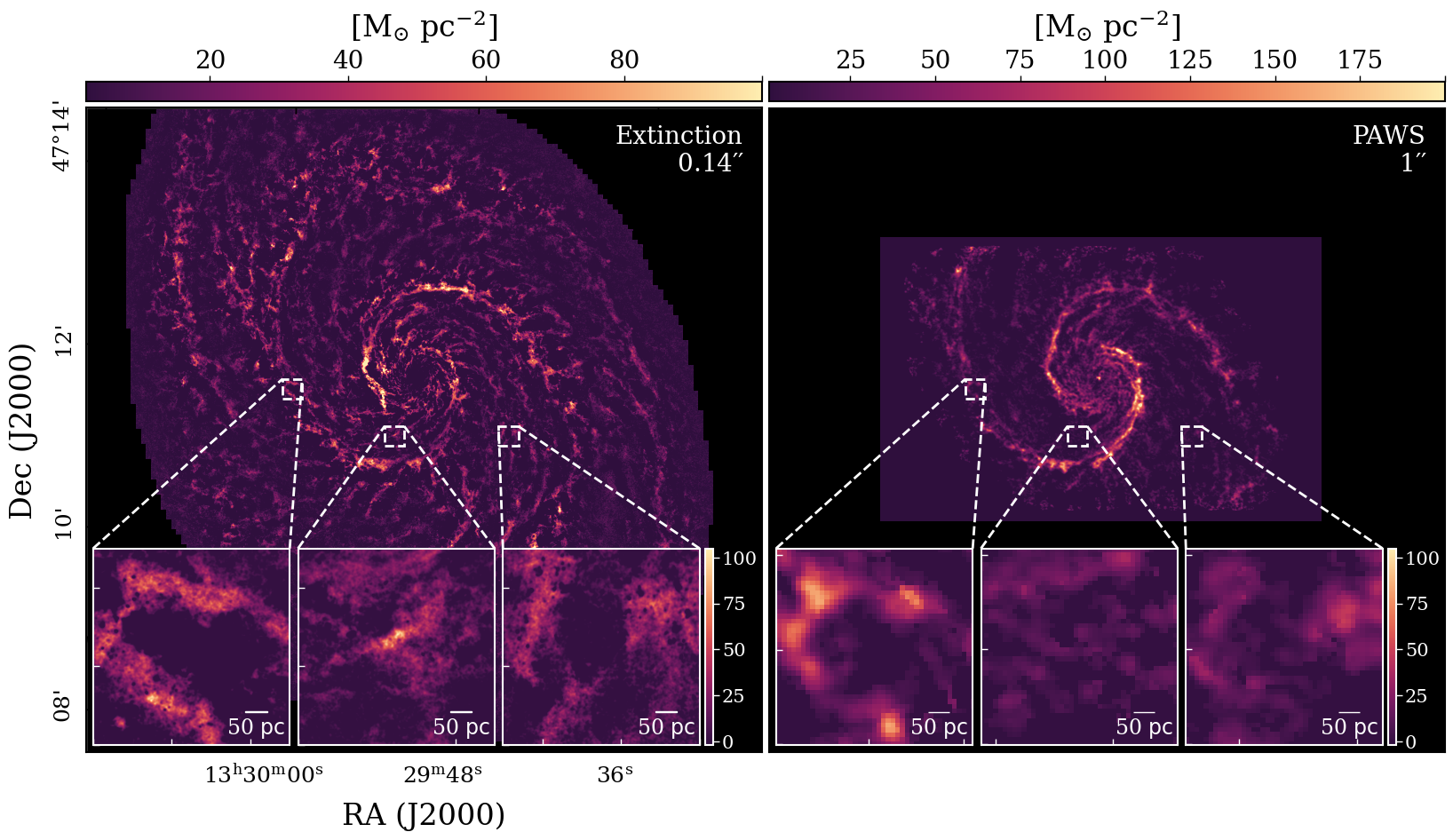}
    \caption{Gas surface density maps, $\Sigma_{\text{gas}}$, from our dust extinction method (\textit{left}) compared to the gas surface densities obtained from converting the PAWS $^{12}$CO(1-0) integrated intensity with ${X_{\text{CO}} = 3.14 \times 10^{19}\,\text{cm}^{-2}\text{(K km s}^{-1})^{-1}}$ (\textit{right}). Zoom-ins of 3 example regions are shown at the bottom of both panels for better visualisation of the spatial resolution achieved by the two different methods (all shown in the same colour scale).}
    \label{fig:2comparison}
\end{figure*}

Molecular line observations of the $^{12}$CO(1-0) emission are one of the most commonly used tracers for the invisible, cold H$_2$ in molecular clouds (MCs). Unlike H$_2$, at the cold temperatures characteristic of MCs, CO - the second most abundant molecule in the ISM - is still easily excited. Despite this, CO emission is not sensitive to the total gas column within a cloud; CO is quickly photodissociated by the interstellar radiation field at low column densities, whilst H$_2$ is still present in these regions due to self-shielding \citep[e.g.][]{lada_near-infrared_2007,penaloza_using_2017,galliano_nearby_2022}. Therefore, surveys that employ CO as a tracer are only sensitive to the "CO-bright" gas, and might miss a significant fraction of cloud masses \citep[e.g.][]{lada_near-infrared_2007, bolatto_co--h2_2013}. Additionally, $^{12}$CO(1-0) is optically thick at relatively low/intermediate column densities and consequently does not trace the full column of gas \citep[see][and references therein]{bolatto_co--h2_2013}. Often, a "CO-to-H$_2$" conversion factor is used to retrieve the total molecular column densities of MCs from CO observations. This conversion factor, $X_{\text{CO}}$, is defined as:
\begin{equation}
    \centering
    \label{eqn:xco}
    X_{\text{CO}} = \text{N(H$_2$)} \, W_{\text{CO}},
\end{equation}

\noindent where $X_{\text{CO}}$ has units ${\text{cm}^{-2}\text{(K km s}^{-1})^{-1}}$, N(H$_2$) is the column density of molecular gas in cm$^{-2}$, and $W_{\text{CO}}$ is the integrated line intensity of $^{12}$CO(1-0) in K km s$^{-1}$. $X_{\text{CO}}$ is an empirically derived value, and is known to have a large uncertainty associated to it \citep[e.g.][]{lada_near-infrared_2007, bolatto_co--h2_2013, barnes_galactic_2018}. The most commonly adopted value of $X_{\text{CO}}$ is the Galactic ${X_{\text{CO}}=2\times10^{20}\,\text{cm}^{-2}\text{(K km s}^{-1})^{-1}}$, obtained through observations of MCs in the Milky Way disc \citep{strong_gradient_1996, dame_milky_2001}. However, many studies suggest that $X_{\text{CO}}$ varies significantly from galaxy to galaxy, and even within different environments of the same galaxy \citep[e.g.][]{pineda_co_2008,bolatto_co--h2_2013, sandstrom_co--h2_2013, barnes_three-mm_2015, gong_environmental_2020}.

In the PAWS field-of-view (FoV, shown in Fig.~\ref{fig:2comparison} alongside our extinction surface densities), \cite{colombo_pdbi_2014} quote $84 \, \sunpc$ as the median gas mass surface density value, derived assuming the Galactic $X_{\text{CO}}$. If we estimate the median in the same region that PAWS imaged (and at the same angular resolution and pixel grid), but using our extinction surface densities instead, we obtain a much lower value ($\sim7 \, \sunpc$). This median value is calculated considering only surface densities within the PAWS FoV above $5 \sigma$, where $\sigma$, the average standard deviation in low-emission areas, is $0.44 \, \sunpc$. Dust extinction traces the total gas, whilst CO traces only the molecular gas; calculating the same median only in regions more likely to be molecular (i.e. $\Sigma > 10 \, \sunpc$) gives a value of roughly $\sim20 \, \sunpc$. Both of the medians we report within the PAWS FoV are much smaller (between a factor 4 to 12 smaller) than the PAWS gas mass surface density median computed using the Galactic $X_{\text{CO}}$ quoted by \cite{colombo_pdbi_2014}, even when only considering pixels above our molecular surface density threshold. 

\cite{guelin_cold_1995} find that their dust-derived molecular masses are about a factor 4 smaller than masses computed from ${^{12}\text{CO(1-0)}}$ emission using the Galactic $X_{\text{CO}}$, suggesting that the CO-to-H$_2$ conversion factor in M51 might be lower than the Galactic value. \cite{bell_molecular_2007} found ${X_{\text{CO}} = 2.5 \times 10^{19}}$ in the centre of M51 through comparison of observed emission line intensity ratios and predictions from photon-dominated region (PDR) chemical models \cite[see also][]{bell_molecular_2006}. Using large velocity gradient (LVG) modelling of several transitions of $^{12}$CO and $^{13}$CO, as well as neutral carbon, \cite{israel_ci_2006} calculated ${X_{\text{CO}} = 1 (\pm 0.5) \times 10^{20}}$ for the centre of M51. On the other hand, several studies reiterate that the standard Galactic $X_{\text{CO}}$ should be applicable in M51 \citep[e.g.][]{schinnerer_multi-transition_2010,leroy_cloud_2017}, given its nearly constant solar metallicity \citep[e.g.][]{croxall_chaos_2015}.

\begin{figure}
    \centering
    \includegraphics[width=0.45\textwidth]{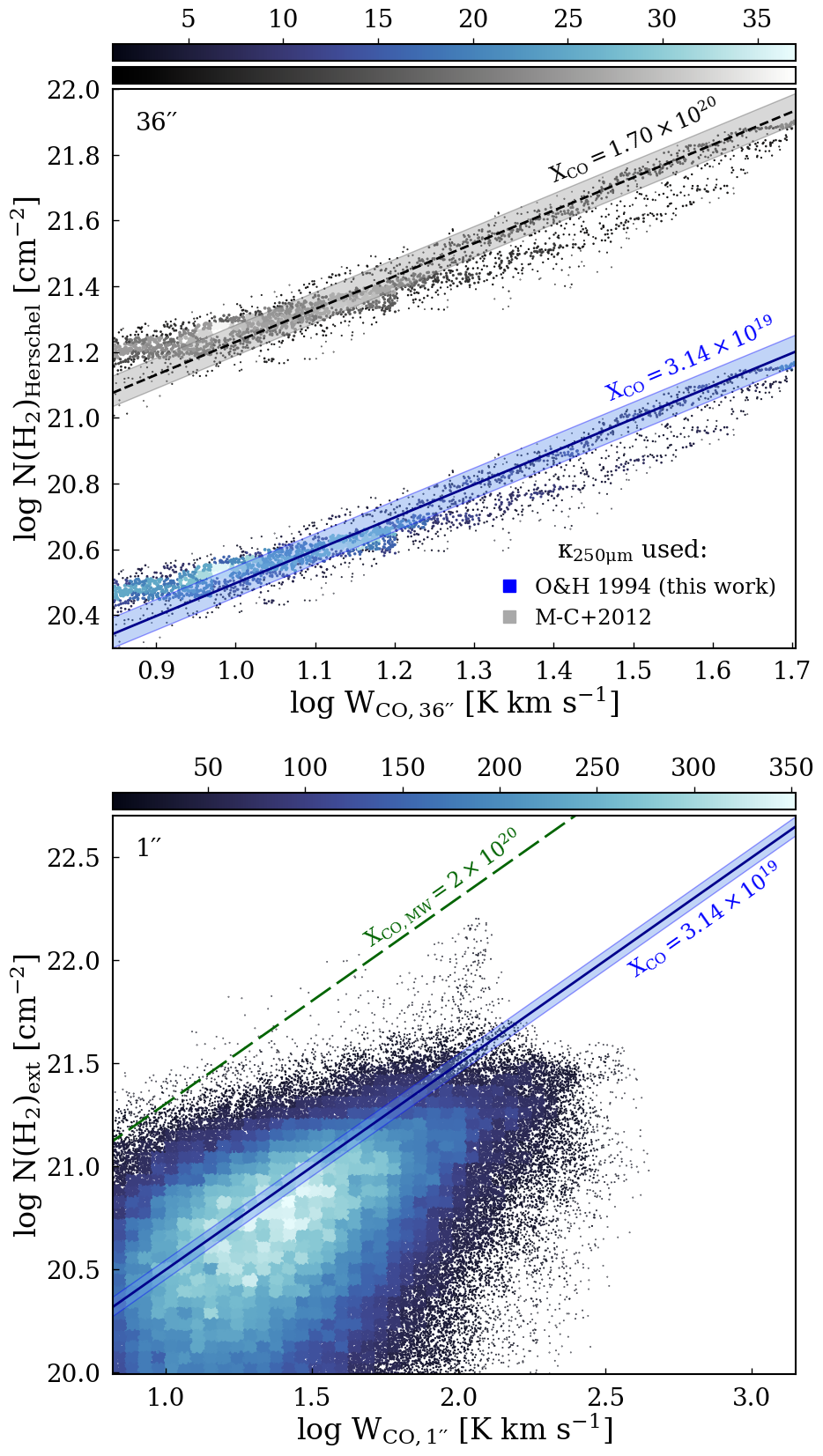}
    \caption{\textit{Top:} H$_2$ column densities from the \textit{Herschel} dust emission data, $\text{N(H}_2\text{)}_{\textit{Herschel}}$, against the integrated line intensity of $^{12}$CO(1-0) from PAWS, $W_{\text{CO}, 36"}$, at the resolution of the dust emission data (36"). The different colours represent the resulting dust emission column densities from adopting two different dust absorption coefficients: the \protect\cite{ossenkopf_dust_1994} specific opacity in blue (used in this work), and the \protect\cite{mentuch_cooper_spatially_2012} specific opacity in grey (see §\ref{sec:differentkappa}). \textit{Bottom:} H$_2$ column densities resulting from our extinction technique, $\text{N(H}_2\text{)}_{\text{ext}}$, also against the integrated line intensity of $^{12}$CO(1-0) from PAWS, at the resolution of PAWS ($W_{\text{CO}, 1"}$). In both panels: the solid blue line represents the value of $X_{\text{CO}}$ we retrieve assuming the \protect\citealt{ossenkopf_dust_1994} (O\&H 1994) dust absorption coefficient in the IR (see text for details) with the blue-shaded region being the interquartile spread on said $X_{\text{CO}}$ at the \textit{Herschel} resolution. The blue (and grey) scale in both plots represents the density of points. In the top panel, the dashed black line represents the resulting $X_{\text{CO}}$ when assuming $\kappa_{250\upmu\text{m}}$ from \protect\citealt{mentuch_cooper_spatially_2012} (M-C+2012). In the bottom panel, the green dashed line depicts the standard Galactic $X_\text{CO,MW}$.}
    \label{fig:both_cd_paws}
\end{figure}

In order to make our extinction method's results comparable to the CO results from PAWS, instead of assuming an ad-hoc value such as the Galactic $X_{\text{CO}}$, we rederive the $X_{\text{CO}}$ by comparing the PAWS $W_{\text{CO}}$ map (§\ref{sec:paws}) directly to the surface densities we computed from the \textit{Herschel} data (§\ref{sec:herschel}), as this is also what we used to calibrate our extinction column densities. We convert the dust emission surface densities into molecular gas column densities, N(H$_2$), through: 

\begin{equation}
    \text{N(H$_2$)} = \frac{\Sigma_{\text{em,gas}}}{\mu \, m_H},
    \label{eq:nh2}
\end{equation}

\noindent where $\Sigma_{\text{em,gas}}$ is the gas surface densities from \textit{Herschel} (converted from dust, $\Sigma_{\text{em}}$, assuming a dust-to-gas mass ratio of $1\%$), $m_H$ is the mass of a hydrogen atom, and $\mu$ the mean molecular weight. We adopt a mean molecular weight of $\mu=2.8$ \citep{kauffmann_mambo_2008}. The PAWS $W_{\text{CO}}$ map is regridded and convolved to the pixel grid ($3.2"$/pix) and resolution ($36"$) of the \textit{Herschel} column densities. The distribution of the \textit{Herschel} N(H$_2$) against the convolved PAWS $W_{\text{CO}}$ is depicted in Fig.~\ref{fig:both_cd_paws}. The resulting $X_{\text{CO}}$ is almost a factor 7 smaller than the Galactic standard: we retrieve a median of ${X_{\text{CO}} \simeq 3.14 \times 10^{19}}$, with lower and upper quartiles of ${X_{\text{CO}} \simeq 2.84 \times 10^{19}}$ and ${X_{\text{CO}} \simeq 3.52 \times 10^{19}}$, respectively. Our derived $X_{\text{CO}}$ is within the range of lower $X_{\text{CO}}$ values reported for M51 in the literature. 

As previously discussed in §\ref{sec:differentkappa}, our determination of an $X_{\text{CO}}$ value for M51 will also be heavily influenced by our assumed dust models \citep[see also][]{sandstrom_co--h2_2013,roman-duval_dust_2014}, as are our final gas surface density estimates. Following Eq.\,(\ref{eq:scaling}), we combine Eq.\,(\ref{eqn:xco}) and Eq.\,(\ref{eq:nh2}) to give a scaling relation between our determined $X_{\text{CO}}$ value and the adopted absorption coefficient for the IR:

\begin{equation}
    X_{\text{CO}}^{\text{new}} = 3.14\times10^{19} \, \left( \frac{21.6 \, \text{cm}^{2}\text{g}^{-1}}{\kappa^{\text{new}}} \right), 
    \label{eqn:scaling_xco}
\end{equation}

\noindent where $X_{\text{CO}}^{\text{new}}$ is the CO-to-H$_2$ conversion factor (in ${\text{cm}^{-2} (\text{K km s}^{-1})^{-1}}$) obtained when assuming a dust absorption coefficient, $\kappa^{\text{new}}$, at $250\,\upmu\text{m}$. In particular, taking $\kappa^{\text{new}} = 3.98\,\text{cm}^2\text{g}^{-1}$ from \cite{mentuch_cooper_spatially_2012}, retrieves a value of $X_{\text{CO}}$ within a factor 0.5 from the standard Galactic value (shown in top panel of Fig. \ref{fig:both_cd_paws}).

Our main goal in determining a CO-to-H$_2$ conversion factor is to make the surface densities from PAWS comparable to our results. In fact, if we compare the PAWS $W_{\text{CO}}$ with our extinction column densities at the PAWS resolution (bottom panel of Fig.~\ref{fig:both_cd_paws}), we can see that the ${X_{\text{CO}}\,\simeq\,3.14 \times\,10^{19}}$\,cm$^{-2}$\,(K\,km\,s$^{-1}$)$^{-1}$ derived from the comparison with \textit{Herschel} dust emission at kpc-scales works reasonably well at the higher resolution. It is also clear from this figure that adopting the Galactic $X_{\text{CO}}$ would overestimate the molecular masses from CO, producing the discrepancy of statistics previously mentioned. Applying the new $X_{\text{CO}}$ to the $W_{\text{CO}}$ from PAWS gives a median mass surface density of $\sim 18 \, \sunpc$ for M51, which is now consistent with the median values we report for the same region at the PAWS resolution. This calculation is only performed in regions with significant CO detections\footnote{The PAWS $^{12}$CO(1-0) map is already masked to only contain high-fidelity CO detections.}, and where our gas surface densities are above $5\sigma$ ($\sigma = 0.44 \, \sunpc$, as previously mentioned).

Comparing the spatial distributions of the gas surface densities from PAWS with those from the extinction method, we find that there is no substantial change in the surface density medians for the spiral arms between the rescaled PAWS surface densities and our surface densities at the same resolution ($\sim 21 \, \sunpc$ and $\sim 22 \, \sunpc$, respectively). There is, however, a difference between the PAWS inter-arm surface density median and our value, with $\sim 12 \, \sunpc$ and $\sim 17 \, \sunpc$, respectively. This discrepancy may be due to the enhanced presence of CO-dark gas towards the inter-arms, where CO does not trace the full column of gas due to the lack of shielding, whereas dust extinction is still sensitive. 

The median from the rescaled PAWS surface densities for the centre of M51 is $37 \, \sunpc$, higher than the median we report for the same area ($23 \, \sunpc$). In the centre of M51 most of the gas is CO-bright, whilst dust extinction is limited to lower estimates of column due to saturation effects, and failures in the removal of bright sources. On the other hand, the centre of M51 is known to host different conditions than the disc \citep[e.g.][]{mentuch_cooper_spatially_2012,schinnerer_pdbi_2013,nersesian_high-resolution_2020}. In particular, the dust in that region is hotter due to considerable heating caused by the dominant old stellar population. Therefore, CO-based measurements will be less accurate due to temperature effects. Furthermore, applying a single value of $X_{\text{CO}}$ for the centre and the disc is likely not a fair assumption under these conditions \citep[see][]{sodroski_ratio_1995, bolatto_co--h2_2013, gong_environmental_2020}.

\subsection{Comparison with B-band surface densities, $\Sigma_{B}$}

The B-band emission is more heavily attenuated by the interstellar dust grains than the V-band, such that we reach saturation with lower dust columns. Consequently, the surface densities of the denser regions of M51 will be less accurately determined with the B-band than with the V-band, and thus we preferably use the V-band for our method. Nonetheless, our technique will still work for the B-band. 

We adopt a dust mass absorption coefficient for the B-band of $\kappa_B=1.19\times10^4$ cm$^2$ g$^{-1}$ ($\kappa_B = 2.4858$ pc$^2$ M$_\odot^{-1}$) from D03. Assuming an initial guess of $b' = f' = 0.5$, our technique corrects the background and foreground fractions of the B-band to $b=0.52$ and $f=0.48$, respectively. The ratio between our calibrated maps of $\tau_V$ and $\tau_B$ has a median of $\sim 0.73$, which is in agreement with the ratio of the respective specific opacity laws ($\kappa_V / \kappa_B \sim 0.72$). The resulting surface densities, $\Sigma_B$, are similar to the final V-band surface densities, with a median $\Sigma_{\text{gas}, V} / \Sigma_B \sim 1.04$, highlighting the fact that the V-band can reach higher surface densities, but that within the uncertainties, both bands perform equally well across the galaxy.

\section{Summary and conclusions}
\label{sec:sum_conc}

%The last numbered section should briefly summarise what has been done, and describe the final conclusions which the authors draw from their work.

Here we present a new dust extinction-based technique that allows the retrieval of high-resolution gas surface density maps for entire galaxies. The method is based on similar work done for our Galaxy in the IR, and consists of determining the dust attenuation pixel by pixel in the optical, through comparison with a reconstructed, smoothly varying stellar background. We applied this method to M51 as a test case, and our final gas surface density map has a resolution of $0.14"$ (or $\sim5$ pc at the adopted distance), which is a factor $\sim 7$ better physical resolution than the currently highest resolution CO dataset for M51 \citep[PAWS,][]{schinnerer_pdbi_2013}. 

We compare our surface density estimates to several independent dust- and CO-based approaches, and find that our map correlates well with lower-resolution dust (and also gas) maps of M51. In particular, we find similar trends of surface density across large-scale environment as those seen in PAWS \citep{colombo_pdbi_2014}. Any disparities in gas/dust surface density estimates between our map and other studies arise primarily from the different dust opacity laws used. If the a priori assumptions of dust model are the same, our technique retrieves values of column consistent with independent studies. We provide a scaling relation applicable to our high-resolution gas surface density map of M51, to account for the use of different dust absorption coefficients in the IR.

We retrieve a CO-to-H$_2$ conversion factor almost a factor 7 lower than the standard Galactic $X_{\text{CO}}$. We find that the choice of absorption coefficient in the modelling of dust emission (i.e. adopting different dust compositions and therefore emissivity) has a significant influence in the determination of $X_{\text{CO}}$. Using the specific opacity law at intermediate surface densities from OH94 results in a conversion factor ${X_{\text{CO}} = 3.1\,(\pm 0.3)\times10^{19}\,\text{cm}^{-2}\,\text{(K km s}^{-1})^{-1}}$, assuming a constant dust-to-gas ratio of $1\%$. Using a lower dust absorption coefficient instead will result in much higher dust masses, which in turn returns a higher $X_{\text{CO}}$ value. In fact, with the absorption coefficient from \cite{draine_infrared_2007}, we obtain a value similar to the Galactic ${X_{\text{CO}}}$ (${X_{\text{CO}}\,=\,1.7\,(\pm 0.2)\times 10^{20}\,\text{cm}^{-2}\,\text{(K km s}^{-1})^{-1}}$).

In summary, the strength of our extinction technique lies in the high spatial resolution of the resulting map (almost an order of magnitude higher than previous studies), obtained from readily available optical data. Furthermore, we are able to probe the lower surface density regime composed of atomic and/or CO-dark molecular gas, and thus providing new tools to study the ISM and its morphology in both its atomic and molecular forms. Our extinction technique is also applicable to the full disc of galaxies, providing a wider coverage which greatly enhances number statistics and completeness. In a follow-up paper (Faustino Vieira in prep), we extract a catalogue of clouds from our gas surface density map of M51, and perform a statistical analysis of resolved cloud populations as a function of large-scale galactic environment as well as galactocentric radius. Our spatially resolved information opens up the door for such studies to be conducted across nearby galaxies of various morphological types, and examine the impact of the different dynamics and environments of galaxies, as well as feedback mechanisms, on the shaping and evolution of their gas content. Additionally, with high-resolution maps such as ours, its is possible to study the spatial coincidence between dust extinction and other H$_2$ tracers (such as CO) as well as SF tracers. In fact, our achieved spatial resolution is on par with the emerging nearby galaxy observations from PHANGS-JWST \citep{lee_phangsjwst_2023}, allowing for direct comparisons between dust extinction and emission.

\section*{Acknowledgements}

%The Acknowledgements section is not numbered. Here you can thank helpful colleagues, acknowledge funding agencies, telescopes and facilities used etc. Try to keep it short.

We thank the anonymous referee for their helpful feedback, which improved the manuscript. HFV and ADC acknowledge the support from the Royal Society University Research Fellowship URF/R1/191609. TAD, NP, MWLS and MA acknowledge support from the UK Science and Technology Facilities Council through grants ST/S00033X/1 and ST/W000830/1. The calculations performed here made use of the computing resources provided by the Royal Society Research Grant RG150741. MQ acknowledges support from the Spanish grant PID2019-106027GA-C44, funded by MCIN/AEI/10.13039/501100011033. HFV would like to thank Sharon Meidt for allowing the use of the PAWS environmental mask, and Chris Clark for useful discussions. Based on observations made with the NASA/ESA Hubble Space Telescope, which is operated by the Association of Universities for Research in Astronomy, Inc. (Program \#10452). DustPedia is a collaborative focused research project supported by the European Union under the Seventh Framework Programme (2007-2013) call (proposal no. 606847). The participating institutions are: Cardiff University, UK; National Observatory of Athens, Greece; Ghent University, Belgium; Université Paris Sud, France; National Institute for Astrophysics, Italy and CEA, France.

\section*{Data availability}
With this paper, we release the extinction-derived gas surface density map of M51 in \url{https://dx.doi.org/10.11570/23.0010} and in the FFOGG (Following the Flow of Gas in Galaxies) project website (\url{https://ffogg.github.io/ffogg.html}).

%%%%%%%%%%%%%%%%%%%%%%%%%%%%%%%%%%%%%%%%%%%%%%%%%%



\bibliographystyle{mnras}
\bibliography{main.bbl} % if your bibtex file is called example.bib

%\begin{thebibliography}{99}
%\end{thebibliography}

%%%%%%%%%%%%%%%%%%%%%%%%%%%%%%%%%%%%%%%%%%%%%%%%%%

%%%%%%%%%%%%%%%%% APPENDICES %%%%%%%%%%%%%%%%%%%%%

\appendix

\section{Uncertainties in our opacity estimates}
\label{sec:appA}

Assuming a fixed specific opacity law, the uncertainty in our opacity estimates (and consequently in our extinction technique) will translate into relative errors for cloud surface densities/masses. In order to quantify the uncertainty in the opacities, $\sigma_{\tau}$, we measure the standard deviation obtained across $10^4$ Monte Carlo realizations of $\tau$ for each pixel.

\begin{figure}
    \centering
    \includegraphics[width=0.4\textwidth]{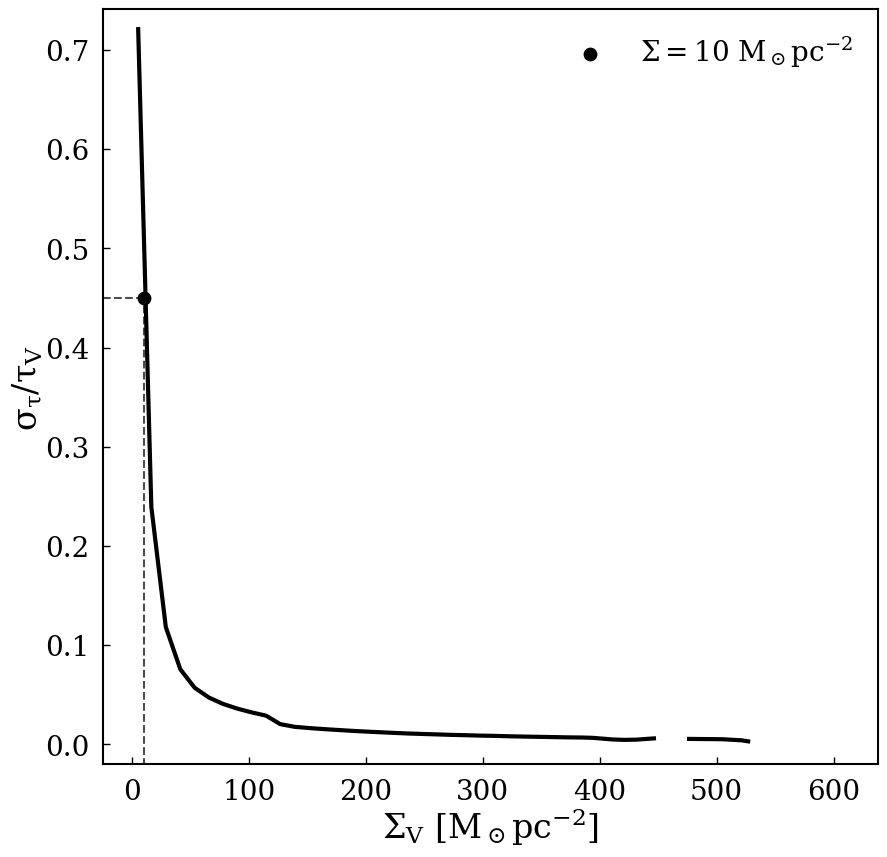}
    \caption{Running median of the relative error on the opacities ($\sigma_\tau/\tau_V$) against our extinction gas surface densities, $\Sigma_V$. Our adopted molecular surface density threshold, $\Sigma > 10 \, \sunpc$, corresponds to a maximum relative uncertainty of 45\% (highlighted in the figure by the solid black dot and dashed lines), although note that it decreases very rapidly to well below 30\% uncertainty.}
    \label{fig:uncertainties}
\end{figure}

To showcase the dependence of $\tau$ in $I_V$, $I_0$, and $b$, we rearrange Eq. (\ref{eqn:method}) as

\begin{equation}
    \label{eqn:MC}
    \tau = - \text{ln} \left( \frac{I_V - (1-b) \, I_0}{b \, I_0} \right),
\end{equation}

\noindent where $I_V$ is the source-removed intensity, $b$ the assumed background light fraction and $I_0$ the reconstructed stellar distribution (see §\ref{sec:method}). In our Monte Carlo simulations we allow these 3 parameters to vary within their respective standard deviations. We estimate $\sigma_{I_V}$ (i.e. the photometric noise) by measuring and averaging the standard deviation within low-emission regions of our source-removed map, and obtained $\sigma_{I_V} = 8.59 \times 10^{-3}$. Given that the stellar distribution is reconstructed by applying a median filter on the source-removed map, we assume that the uncertainty in $I_0$, $\sigma_{I_0}$, will be related to the photometric noise through: $\sigma_{I_0} = 1.2533 \, \sigma_{I_V} \, \sqrt{\text{MF}} = 5.94 \times 10^{-4}$, i.e. the standard error on the median, where MF is the median filter used in pixels (330 pix, see §\ref{sec:stellardist}). We take the scatter on our estimate of $b$ (see §\ref{sec:calibration}) as its uncertainty, resulting in $\sigma_{b} = 6.65 \times 10^{-3}$. The running median of the relative error in the opacities (i.e. $\sigma_{\tau}/ \tau$) is shown in Fig.~\ref{fig:uncertainties}, as a function of the local surface density $\Sigma_V$. Above $10\,\sunpc$ (the often quoted molecular threshold), the maximum relative uncertainty is 45\%, rapidly dropping below 30\% above $14\,\sunpc$.

Besides these relative uncertainties on the opacities (and masses/surface densities), there is another key observational limitation to our method, which is the fact that the amount of extinction we can effectively measure depends not only on the photometric noise, but also on the background level. Indeed we expect that the maximum opacity we can measure will be lower for lower intensity backgrounds. Thus we need to ensure that our measured surface densities are well below our maximum measurable limit. 

\begin{figure}
    \centering
    \includegraphics[width=0.4\textwidth]{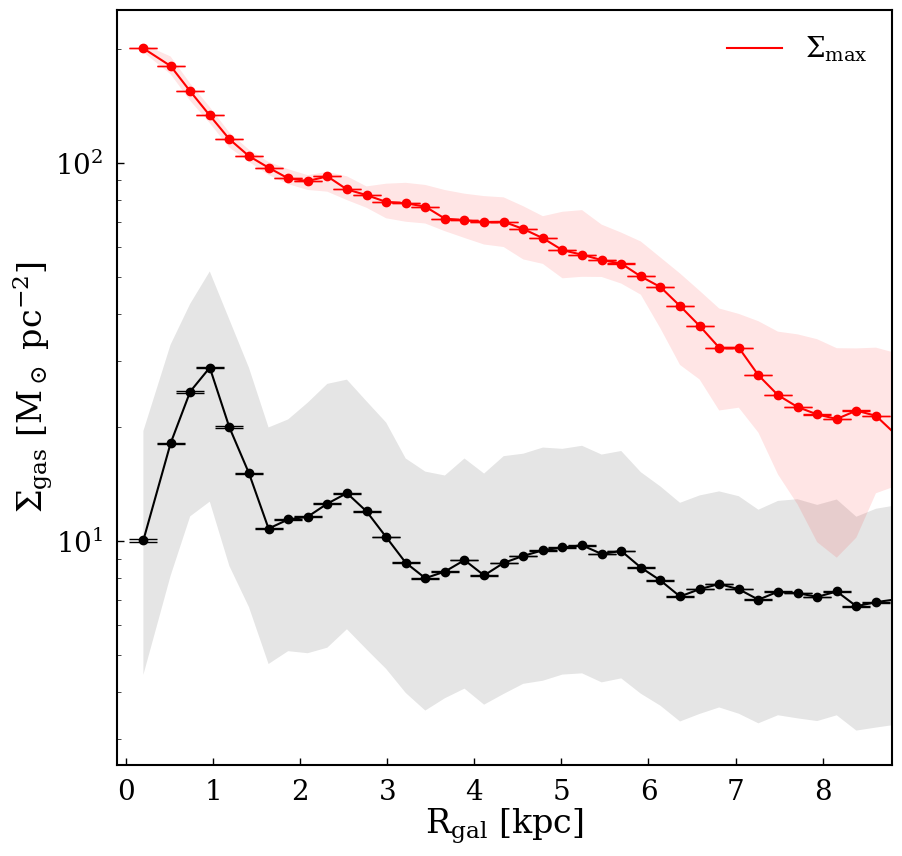}
    \caption{Comparison between the running median of our gas surface densities, $\Sigma_{\text{gas}}$ (black line and dots), and the running median of the maximum value of surface density we can accurately measure, $\Sigma_{\text{max}}$, given the photometric uncertainties in our HST data (red line and dots), as a function of galactocentric radius, $R_{\text{gal}}$. The errorbars depict the standard error on the median, defined as $1.253 \, \sigma/\sqrt{N}$ ($N$ being the bin count and $\sigma$ the standard deviation). The black and red shaded regions represent the interquartile ranges for $\Sigma_{\text{gas}}$ and $\Sigma_{\text{max}}$, respectively.}
    \label{fig:sdmax}
\end{figure}

In order to verify this, we determine the maximum opacity, $\tau_{\text{max}}$, (and consequently surface density) that we are able to reliably measure with our extinction technique at each pixel. 
Remembering Eq.\,(\ref{eqn:method}), we obtain our maximum value of $\tau$ when $I_V - I_{\text{fg}}$ is at a minimum. If we impose that ${I_V - I_{\text{fg}} = 3\sigma_{I_V}}$, i.e. if this minimum cannot be lower than 3 times the photometric error of our star-subtracted image, $\sigma_{I_V}$, we can write that

\begin{equation}
    \tau_{\text{max}} = - \text{ln} \left( \frac{3 \sigma_{I_V}}{I_{\text{bg}}} \right),
    \label{eqn:taumax}
\end{equation}

\noindent where $I_{\text{bg}}$ is the background fraction of the stellar distribution such that $I_{\text{bg}} = b \, I_0$ ($b = 0.53$, see §\ref{sec:calibration}). The maximum surface density we can thus measure, $\Sigma_{\text{max}}$, is related to $\tau_{\text{max}}$ by $\Sigma_{\text{max}} = \tau_{\text{max}} / \kappa_V$. Figure\,\ref{fig:sdmax} shows the running median of $\Sigma_{\text{max}}$ as a function of galactocentric radius. It is clear from the figure that, on average, our extinction gas surface densities are below the maximum surface density we are able to measure. 

\section{Temperature effects in the centre of M51}
\label{sec:appC}

When calibrating our dust extinction surface densities, $\Sigma_{\text{ext}}$, with dust emission, $\Sigma_{\text{em}}$, we opted to exclude the contribution from pixels with large $\Sigma_{\text{em}}$ located at the centre of M51 (see §\ref{sec:calibration} and Fig.\,\ref{fig:linear_fit}), as those are the points at which the $\Sigma_{\text{em}}$ and $\Sigma_{\text{ext}}$ stop following a linear trend, potentially due to temperature effects (or different dust properties).

\begin{figure}
    \centering
    \includegraphics[width=0.4\textwidth]{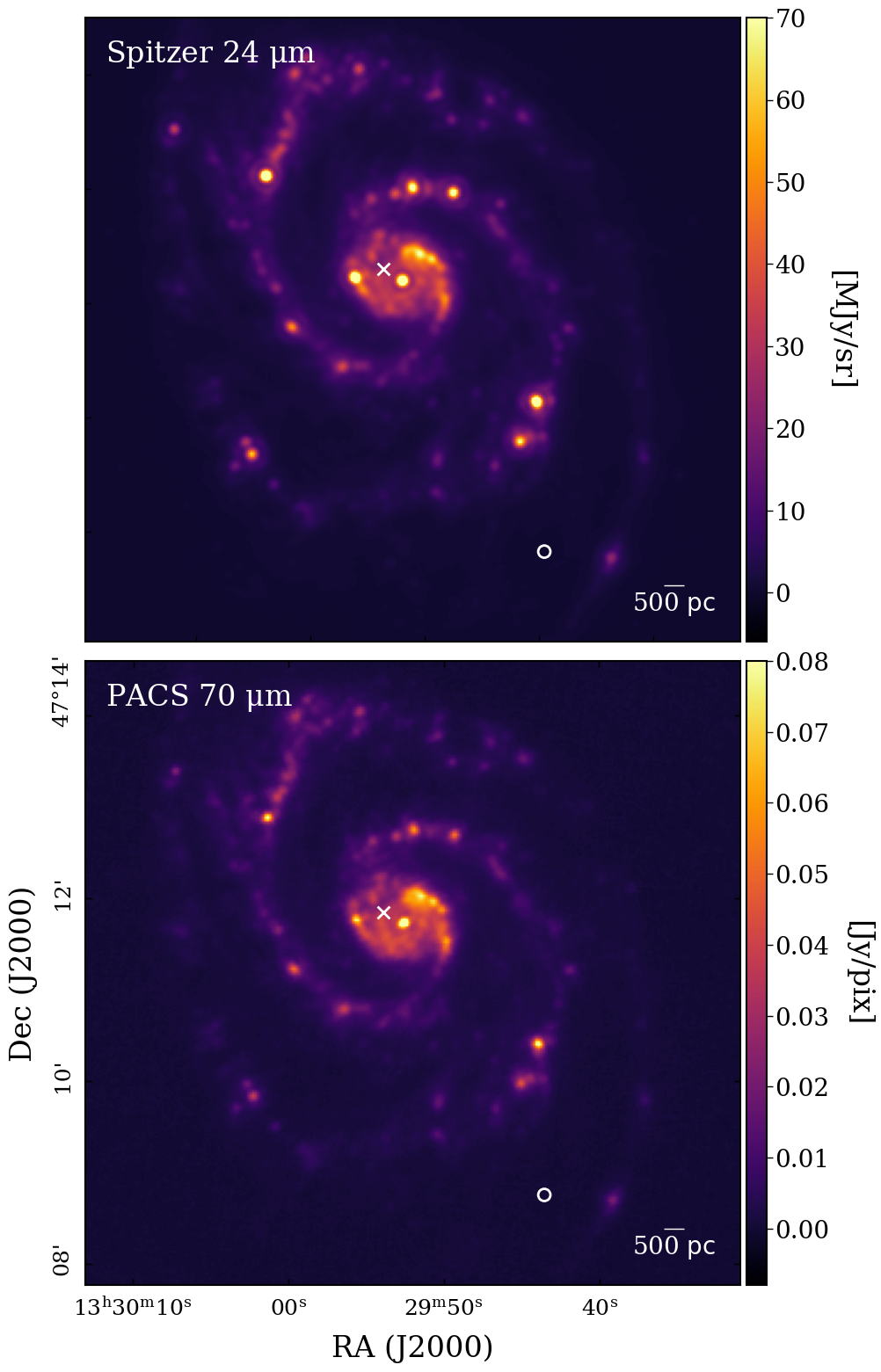}
    \caption{Zoom-in of the $24 \, \upmu$m \textit{Spitzer} MIPS (\textit{top}) and the $70 \, \upmu$m \textit{Herschel} PACS (\textit{bottom}) observations of M51. The centre of the galaxy hosts extended warm emission, likely due to heating from old stars \citep[e.g.][]{schinnerer_pdbi_2013}. The white cross and white circle depict two representative pixels from the centre and disc of the galaxy, respectively.}
    \label{fig:2470}
\end{figure}

Unlike dust extinction, dust emission is heavily dependent on the dust temperature along the line-of-sight. As we employ a simple single-temperature modified blackbody model to retrieve measurements of column density and temperature for \textit{Herschel} dust emission observations, we intrinsically assume that dust properties do not vary significantly not only along the line-of-sight but also across M51. In the centre of M51 ($R_{\text{gal}} \lesssim 2$ kpc), where a multitude of physical processes are in play, this assumption may not hold. In fact, \cite{munoz-mateos_radial_2009} report an excess of dust luminosity in relation to the derived dust mass for the inner 2 kpc of the galaxy. From Fig.~\ref{fig:2470}, we can see that in the centre of M51 the dust emission is dominated by extended emission in the MIPS $24 \, \upmu$m (Multiband Imaging Photometer for \textit{Spitzer}, \citealt{rieke_multiband_2004}) and PACS $70 \, \upmu$m dust maps. An increase of $24 \, \upmu$m and $70 \, \upmu$m emission indicate an increase of dust temperatures, which is indeed what is reported by \cite{mentuch_cooper_spatially_2012} in M51's centre. This increase in luminosity (and consequently temperature) can be caused by heating from the prominent old stellar population observed at small galactocentric radii \citep[e.g.][and references therein]{schinnerer_pdbi_2013, nersesian_high-resolution_2020}. 

\begin{figure}
    \centering
    \includegraphics[width=0.4\textwidth]{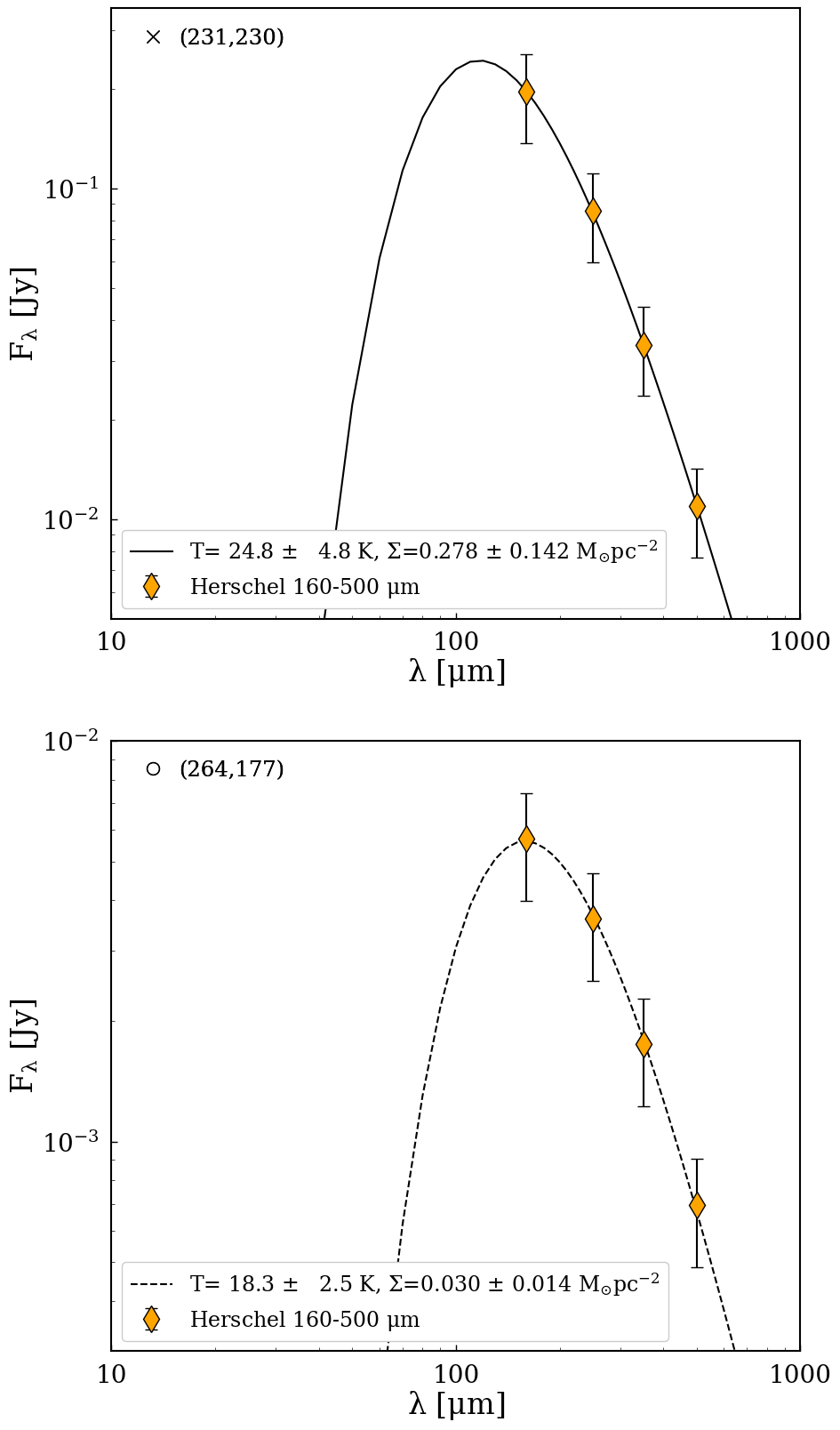}
    \caption{Examples of fitted SED curves for a representative pixel within the centre of M51 (\textit{top}), and for a pixel within the disc of the galaxy (\textit{bottom}). In both plots, the upper left corner shows the coordinates of the chosen pixel, as well as the corresponding marker shown in Fig.~\ref{fig:2470}. The orange diamonds represent the fitted \textit{Herschel} data points, with the black errorbars representing the uncertainty on the flux. For the central pixel shown in the top panel, our modified blackbody model fits a temperature of $T = 25 \pm 5$ K and a dust surface density of $\Sigma = 0.28 \pm 0.14 \, \sunpc$. For the disc pixel in the bottom panel, we retrieve $T = 18 \pm 2$\,K and $\Sigma = 0.03 \pm 0.01 \, \sunpc$.}
    \label{fig:sedcurve}
\end{figure}

Our single-temperature SED fit from 160\,$\upmu$m onwards cannot accurately describe this diffuse warm emission in the centre. Figure \ref{fig:sedcurve} depicts an example of an SED curve of a pixel within the inner 2 kpc of M51 (highlighted by a white cross in Fig.~\ref{fig:2470}). It is clear that with our wavelength coverage we are not able to effectively capture the turnover in the SED towards shorter wavelengths, which is reflected on the larger uncertainties in the fitted temperatures and column densities. The bottom panel of Fig. \ref{fig:sedcurve} shows instead the SED curve of a pixel within the disc (highlighted by a white circle in Fig.~\ref{fig:2470}), where the SED turnover is better constrained. In order to capture the higher temperatures in the centre, we would require a more sophisticated SED model to fit \textit{Spitzer} and \textit{Herschel} emission together (rather than a simple blackbody curve), which is beyond the scope of this paper. Since our goal was simply to retrieve a statistically reliable conversion between dust emission and dust extinction column densities, we opt for using the bulk of the disk of M51 where our simplified SED fitting is more reliable.

%%%%%%%%%%%%%%%%%%%%%%%%%%%%%%%%%%%%%%%%%%%%%%%%%%

% Don't change these lines
\bsp	% typesetting comment
\label{lastpage}
\end{document}